\documentclass[12pt,letterpaper]{article}
\pdfoutput=1
\usepackage{jheppub}
\usepackage{dsfont}
\usepackage{amssymb,amsmath}
\usepackage{epsfig}
\usepackage{epstopdf}
\usepackage{latexsym}
\usepackage{graphicx}
\usepackage{subfigure}
\usepackage{booktabs}
\usepackage{bbm}
\usepackage[multiple]{footmisc}
\makeatletter
\def\@fpheader{\relax}
\makeatother

\newcommand{\cH}{{\mathcal{H}}}

\newcommand{\be}{\begin{equation}}
\newcommand{\ee}{\end{equation}}
\newcommand{\bea}{\begin{eqnarray}}
\newcommand{\eea}{\end{eqnarray}}

\def\O{{\cal O}}

\def\N{{\cal N}}

\def\S{{\cal S}}
\def\A{{\cal A}}

\def\L{{\cal L}}

\subheader{\begin{flushright}
UTTG-34-13\\
TCC-028-13
\end{flushright}}

\title{Entanglement and out-of-equilibrium dynamics in holographic models of de Sitter QFTs}
\author{Willy Fischler, Sandipan Kundu and Juan F. Pedraza}
\affiliation{Theory Group, Department of Physics and Texas Cosmology Center,\\The University of Texas at Austin, Austin, TX 78712}
\emailAdd{fischler@physics.utexas.edu}
\emailAdd{sandyk@physics.utexas.edu}
\emailAdd{jpedraza@physics.utexas.edu}

\abstract{In this paper we study various aspects of entanglement entropy in strongly-coupled de Sitter quantum field theories in various dimensions. We focus on gravity solutions that are dual to field theories in a fixed de Sitter background, both in equilibrium and out-of-equilibrium configurations. The latter corresponds to the Vaidya generalization of the AdS black hole solutions with hyperbolic topology. We compute analytically the entanglement entropy of spherical regions and show that there is a transition when the sphere is as big as  the horizon. We also explore thermalization in time-dependent situations in which  the system evolves from a non-equilibrium state to the Bunch-Davies state. We find that the saturation time is equal to the light-crossing time of the sphere. This behavior is faster than random walk and suggests the existence of free light-like degrees of freedom.}

\begin{document}

\maketitle
\flushbottom

\section{Introduction}

Quantum field theory in curved spacetimes \cite{bd,Wald:1995yp} is a subject of great
relevance that has lead to many interesting areas of research in the past few decades. Although it is believed to be a good physical description in circumstances
where quantum gravitational effects do not play a dominant role, the reader must also be aware of its limitations. For instance, current research suggests that effective QFT might break down in places where the curvature of spacetime is not so large, 
leading to various well-known puzzles and paradoxes \cite{Almheiri:2012rt,Almheiri:2013hfa,braunstein}. Albeit gravity is treated classically, QFT in a fixed background has provided us historically with a tool to investigate qualitatively some of the most fundamental questions in quantum gravity, \emph{e.g.} understanding various aspects of black hole thermodynamics and deriving the physical consequences of inflation and modern cosmology.

QFT in curved spacetimes is often discussed in terms of free field theory but it has been extended to perturbative studies of weakly coupled field theories. While most of the qualitative features are visible at this level, understanding the strong-coupling and non-perturbative dynamics is of interest. Indeed, in the context of cosmology it is plausible that strong dynamics might have played an important role in the early universe.

In recent years, the discovery of the AdS/CFT correspondence
\cite{Maldacena:1997re,Gubser:1998bc,Witten:1998qj} has
provided tools for the study of a large class of strongly-coupled QFTs. To date, there have been a number of proposals for
holographic theories living in de Sitter space and other cosmological backgrounds \cite{Buchel:2002wf,Buchel:2002kj,Aharony:2002cx,Balasubramanian:2002am,Cai:2002mr,Ross:2004cb,Alishahiha:2004md,Balasubramanian:2005bg,Alishahiha:2005dj,Buchel:2006em,Hirayama:2006jn,Ghoroku:2006af,He:2007ji,Ghoroku:2006nh,Hutasoit:2009xy,Marolf:2010tg,Li:2011bt,Erdmenger:2011sy,Ghoroku:2012vi,Buchel:2013dla}.
The purpose of the present paper is to study entanglement generated during the cosmological evolution \cite{Hawking:2000da,Iwashita:2006zj,MartinMartinez:2012sg,Maldacena:2012xp,Feng:2012km,Adamek:2013vw,Engelhardt:2013jda,Lello:2013qza,Takook:2013boa,Hu:2013ypa}. Entanglement entropy (or geometric entropy) is an important concept and a useful tool in quantum field theories and quantum many body systems, and serves as a probe to characterize states of matter with long range correlations. As a first step we will consider strongly coupled gauge theories in de Sitter spacetime in the large-$N$ limit and we will analytically compute entanglement entropies of spherical regions in the conformally flat patch and the static patch using the holographic prescription.  We will then use entanglement entropy as a tool to study thermalization of out of equilibrium configurations in de Sitter space.

It is well known that entanglement entropy of a spatial region in a local field theory is UV-divergent
\begin{equation}
S=S_{div}+S_{finite} \nonumber\ .
\end{equation}
The remarkable feature of the entanglement entropy in de Sitter is that $S_{div}$  is fully regulated by subtracting the flat space result.\footnote{It is also known that the divergent piece of the entanglement entropy at finite temperature in the flat space-time can also be regulated by subtracting the zero temperature result.} On the other hand, $S_{finite}$ contains information about the long range entanglement and it is expected to be more sensitive to the curvature of  space-time. Another local way to deal with the UV divergences was introduced  in \cite{Liu:2012eea} which amounts to compare entanglement entropies of spheres of similar radii.

Let us now focus on the finite part $S_{finite}$. When the size of the sphere is much smaller than the de Sitter horizon, $R<<H^{-1}$, $S_{finite}$ is expected to be the same as in flat space-time. However, for a particular conformal field theory\footnote{Note that in de Sitter there are gravitational conformal anomalies in even space-time dimensions.} (CFT) which has a gravity dual we will show that $S_{finite}$ in $(1+1)$ and $(2+1)$ dimensions is exactly the same as in flat space-time for $R<H^{-1}$.

In $(3+1)$-dimensions, $S_{finite}$ is more sensitive to the curvature and we find for $R<H^{-1}$
\begin{equation}
S_H(R)-S_0(R)=-c ~A H^2\ , \nonumber
\end{equation}
where $c$ is a constant and $A$ is the proper area of the sphere. We will also show that the entanglement entropy of a sphere of radius $R$ in the static patch of de Sitter is the same as the entanglement entropy of a sphere of proper radius $R$ in the conformally flat patch of de Sitter when the size of the sphere is smaller than the size of the event horizon. Behaviors of the extremal surfaces for radii smaller and larger than the horizon are entirely different and as a consequence the entanglement entropy undergoes a {\it phase transition} at $R=H^{-1}$. This phase transition is a signal of a drastic change in correlations at distance $R=H^{-1}$. It is important to note that  only a Ò{\it super-observer}Ó can ``see" this phase transition of the entanglement entropy.

The above feature is more prominent in the so-called renormalized entanglement entropy, introduced in \cite{Liu:2012eea}, which is a derived quantity that has some advantages over entanglement entropy. Entanglement entropy of a region of size $R$ is sensitive to the physics from scale $R$ all the way down to the short distance cut-off $\epsilon$, whereas renormalized entanglement entropy $\S_d(R)$ is expected to be most sensitive to degrees of freedom at scale $R$ and it somewhat naturally describes the RG flow of the entanglement entropy. In particular, for the vacuum of Lorentz invariant, unitary QFTs $\S_d(R)$ in $(1+1)$ and $(2+1)$ dimensions is monotonically decreasing and non-negative, providing a central function out of the entanglement entropy. In de Sitter we will show that $\S_d(R)=\text{constant}$ when $R<H^{-1}$. In $(1+1)$ and $(2+1)$ dimensions $\S_d(R)=0$ when $R>H^{-1}$ because regions separated by a distance $R>1/H$ are causally disconnected. In $(3+1)$-dimensions, the renormalized entanglement entropy is more complicated and it is neither monotonic nor positive for $R>H^{-1}$. It is  perhaps an indication that the definition of $\S_4(R)$ should be modified in order to construct a central function out of the entanglement entropy.

Entanglement is also a useful probe in out-of-equilibrium configurations. For instance it has been shown that, in comparison to other non-local observables (two-point functions and Wilson loops), entanglement entropy equilibrates the latest when the system undergoes a global quench \cite{AbajoArrastia:2010yt,Balasubramanian:2010ce,Balasubramanian:2011ur}, thus setting the relevant time-scale for the approach to thermal equilibrium. We will explore the issue of thermalization in dS space and we will use entanglement entropy to characterize the time evolution of the system. The time-dependent configurations relevant  in the present context are achieved by turning on for a short time interval a uniform density of sources. The relevant interval of time $\delta t$ is taken to be much
smaller than any other scale in the system, and then is turned
off. The work done by the sources take the system to an
excited state which subsequently equilibrates under the
evolution of the same Hamiltonian before the quench.

In the context of QFT, the study of such non-equilibrium situations is
a serious challenge and a topic of current research. In a recent paper \cite{Calabrese}, Calabrese and Cardy showed that for a variety of $(1+1)$-dimensional CFTs as well as for some lattice models, entanglement entropy for a segment of length $\ell$ grows
linearly in time as
\be
\Delta S(t) \sim \frac{t}{t_{\text{sat}}} S_{\text{sat}}, \nonumber
\ee
and then reaches saturation at some $t=t_{\text{sat}}$. Here, $\Delta S$ represents the
difference of the entanglement entropy with respect to the initial state, and $S_{\text{sat}}$ is the equilibrium value after it reaches saturation. They argued that this remarkably simple behavior
can be understood from a simple model of entanglement
propagation using free-streaming quasiparticles
traveling at the speed of light.

This kind of non-equilibrium configurations has also been considered holographically by studying the gravitational collapse of a shell of null dust in AdS, \emph{i.e.} by means of the so-called Vaidya geometries, starting with the seminal works \cite{Danielsson:1999zt,Danielsson:1999fa,Giddings:2001ii} and continuing with a large body of work that includes the recent additions \cite{Albash:2010mv,Ebrahim:2010ra,Aparicio:2011zy,Balasubramanian:2011at,Allais:2011ys,Keranen:2011xs,Galante:2012pv,Caceres:2012em,Baier:2012tc,Erdmenger:2012xu,Baier:2012ax,Steineder:2012si,Arefeva:2012jp,Bernamonti:2012xv,Baron:2012fv,Caceres:2012px,Balasubramanian:2012tu,Hubeny:2013hz,Steineder:2013ana,Baron:2013cya,Aref'eva:2013wma,Zeng:2013mca,Liu:2013iza,Stricker:2013lma,Balasubramanian:2013oga,Li:2013cja,Zeng:2013fsa,Liu:2013qca}.
Of particular relevance for the present context are the results of \cite{Liu:2013iza,Liu:2013qca}. In these papers Liu and Suh showed that, for holographic theories with a gravity dual, entanglement entropy also undergoes a series of regimes resembling those in phase transitions: pre-local-equilibration quadratic growth, post-local-equilibration linear growth, memory loss, and saturation. These results apply for strongly-coupled QFTs in Minkowski space.

A natural question that arises here is that if these universal features also show up in theories living in different backgrounds, and in de Sitter space in particular. The study of out-of-equilibrium physics in the theories we are considering in the present paper is also interesting on its own right. On one hand, it is known that the early universe went through a series of epochs with varying temperature and intricate strongly-coupled dynamics. While there are known techniques in field theory that allows us to handle certain non-equilibrium problems \cite{Boyanovsky:1993xf,Danielsson:2003wb,Fukuma:2013uxa}, the fact that the theories under consideration are strongly-coupled render the perturbative methods unreliable. Here is where holography plays a useful role.

This rest of the paper is organized as follows: in section \ref{sec2}, we study gravity solutions that are dual to field theories in a fixed de Sitter background. We show explicitly two possible slicings, one that is dual to the static patch and a second one that is dual to the conformally flat patch. In section \ref{sec3} we give a brief review of entanglement entropy, the Ryu-Takayanagi prescription and two useful derived quantities that we shall study in the present context: mutual information and renormalized entanglement entropy. In section \ref{sec4} and \ref{sec5} we compute the entanglement entropy in the Bunch-Davies vacuum of de Sitter QFTs in various dimensions.
Section \ref{sec6} is devoted to study the evolution of entanglement entropy in time-dependent configurations.  Finally, in section \ref{sec7} we make some comments about our results and close with conclusions.

\section{Gravity solutions with dS slices\label{sec2}}

The purpose of this section is to study gravity solutions to Einstein's equations with a foliation such that the boundary metric corresponds to de Sitter space in a given coordinate system. In $(d+1)$-dimensions, the Einstein-Hilbert action with negative cosmological constant is given by
\be\label{action0}
S = \frac{1}{2\kappa^2} \int d^{d+1} x \sqrt{-g} \left(R - 2 \Lambda \right),
\ee
which gives the following equations of motion
\be\label{Eineom}
R_{\mu\nu} - \frac{1}{2} \left(R- 2 \Lambda \right) g_{\mu\nu}=0.
\ee
Here $\kappa^2=8\pi G_N^{(d+1)}$ and $\Lambda=-d(d-1)/2L^2$.

Any asymptotically AdS metric can be written in the Fefferman-Graham form \cite{fg}
\begin{equation}\label{feffermangraham}
ds^2={L^2 \over z^2}\left(g_{\mu\nu}(z,x)dx^{\mu}dx^{\nu}+dz^2\right)~,
\end{equation}
from which the dual CFT metric $ds^2=g_{\mu\nu}(x) dx^{\mu}dx^{\nu}$ can be directly read off as $g_{\mu\nu}(x)= g_{\mu\nu}(0,x)$. The full function $g_{\mu\nu}(z,x)$ also encodes data dual to the expectation value of the CFT stress-energy tensor $T_{\mu\nu}(x)$. More specifically, in terms of the near-boundary expansion
\begin{equation}\label{metricexpansion}
g_{\mu\nu}(z,x)=g_{\mu\nu}(x)+z^2 g^{(2)}_{\mu\nu}(x)+\ldots +z^d g^{(d)}_{\mu\nu}(x)+ z^d \log (z^2) h^{(d)}_{\mu\nu}(x)+\ldots~,
\end{equation}
the standard GKPW recipe for correlation functions \cite{Gubser:1998bc,Witten:1998qj} leads after appropriate holographic renormalization to \cite{dhss,skenderis,skenderis2}
\begin{equation}\label{graltmunu}
\left\langle T_{\mu\nu}(x)\right\rangle={ d\, L^{d-1} \over 16\pi G^{(d+1)}_{N}}\left(g^{(d)}_{\mu\nu}(x)+X^{(d)}_{\mu\nu}(x)\right)~,
\end{equation}
where $X^{(d)}_{\mu\nu}=0$ $\forall$ odd $d$ (reflecting the fact that for odd boundary dimensions, there are no gravitational conformal anomalies),
\begin{eqnarray}\label{xmunu}
X^{(2)}_{\mu\nu}&=&-g_{\mu\nu}g^{(2)\alpha}_{\alpha}~,\\
X^{(4)}_{\mu\nu}&=&-{1\over 8}g_{\mu\nu}\left[\left(g_{\alpha}^{(2)\alpha}\right)^2
-g_{\alpha}^{(2)\beta}g_{\beta}^{(2)\alpha}\right]
-{1\over 2}g_{\mu}^{(2)\alpha}g_{\alpha\nu}^{(2)}
+{1\over 4}g^{(2)}_{\mu\nu}g_{\alpha}^{(2)\alpha}~,\nonumber
\end{eqnarray}
and $X^{(2d)}_{\mu\nu}$ for $d\geq3$ given by similar but longer expressions that we will not transcribe here. In (\ref{xmunu}) it is understood that the indices of the tensors $g^{(n)}_{\mu\nu}(x)$ are raised with the inverse boundary metric $g^{\mu\nu}(x)$.

\subsection{The static patch}

The de Sitter spacetime in $d$-dimensions has an isometry group of $SO(d,1)$. For free field theory, there is a
family of de-Sitter invariant vacuum states and it is known as the $\alpha$-vacua \cite{Mottola:1984ar,Allen:1985ux}. However, among these states only the Bunch-Davies (or Euclidean) vacuum \cite{Bunch:1978yq} reduces to the standard Minkowski vacuum in the limit $H\to0$.

This state is well defined on the entire manifold but, for concreteness, in this section we will center our discussion in the static patch of de Sitter, which covers the causal diamond associated with a single geodesic observer,
\be
ds^2=-(1-H^2r^2)dt^2+\frac{dr^2}{1-H^2r^2}+r^2d\Omega^2_{d-2}.
\ee
The name ``static'' comes from the fact that it has a killing vector $\partial_t$ associated with the isometry of time translations. Therefore, energy as well as entropy are well defined quantities. For such an observer, the Bunch-Davies vacuum is characterized by a temperature $T_{\text{dS}}=H/2\pi$ that is associated to the presence of a cosmological horizon at $r=1/H$, where $H$ denotes the value of the Hubble constant \cite{Gibbons:1977mu}. This can be obtained by continuing to Euclidian space and imposing regularity on the horizon.

To obtain a solution dual to a QFT in a given background we start by writing the $d+1$-dimensional metric of the bulk in the Fefferman-Graham form. In particular, for static dS we can assume that the bulk is independent of time,
\be
ds^2= \frac{L^2}{z^2}\left(-f(r,z)dt^2+j(r,z)dr^2 + h(r,z) d\Omega^2_{d-2}+dz^2 \right). \label{metric1}
\ee
The next step is to write the functions $f(r,z)$, $j(r,z)$ and $h(r,z)$ as a series expansion in $z$ and solve order by order the Einstein equations (\ref{Eineom}).
\bea
f(r,z)&=& f_0(r)+f_2(r)z^2+f_4(r)z^4+\cdots, \nonumber \\
j(r,z)&=& j_0(r)+j_2(r)z^2+j_4(r)z^4+\cdots,  \\
h(r,z)&=& h_0(r)+h_2(r)z^2+h_4(r)z^4+\cdots. \nonumber
\eea
The first coefficients in the above expansions are related to the boundary metric, so if we want the boundary theory to be defined on the static patch of de Sitter we have to impose that
\be \label{ansatz}
f_0(r)= 1-H^2r^2,\qquad j_0(r)= \frac{1}{1-H^2r^2}, \quad \text{and} \quad h_0(r)=r^2.
\ee

After plugging the ansatz (\ref{ansatz}) in Einstein equations (\ref{Eineom}), we find that for all $d$ the solutions are truncated at order $\mathcal{O}(z^4)$,
\bea
f(r,z)&=& (1-H^2r^2)\left(1-\frac{H^2z^2}{4}\right)^2  , \nonumber \\
j(r,z)&=& \frac{1}{1-H^2r^2}\left(1-\frac{H^2z^2}{4}\right)^2, \label{ansatz-sol} \\
h(r,z)&=& r^2\left(1-\frac{H^2z^2}{4}\right)^2. \nonumber
\eea

For these metrics, the curvature scalars are given by
\bea
R^{\mu\nu\sigma\rho}R_{\mu\nu\sigma\rho}&=&\frac{2d(d+1)}{L^4},\label{R1}\\
R^{\mu\nu}R_{\mu\nu}&=&\frac{d^2(d+1)}{L^4},\\
R&=&-\frac{d(d+1)}{L^2},\label{R3}
\eea
which show that these backgrounds are completely smooth and absent of singularities. The solutions have a regular Killing horizon at $z=2/H$ with constant surface gravity, and this is in fact related to the temperature of the field theory $T_{\text{dS}}=H/2\pi$. In addition, there is also the expected horizon at $r=1/H$ $\forall$ $z$.

It is interesting to note that the authors of \cite{Marolf:2010tg} derived a family of solutions dual to QFTs in de Sitter space for arbitrary temperature (not necessarily on the de Sitter invariant vacuum). These solutions are found to be related by a bulk diffeomorphism (that acts as a boundary conformal transformation) to the so-called hyperbolic (or topological) black holes described in \cite{Emparan:1998he,Birmingham:1998nr,Emparan:1999gf}. We will come back to these solutions in section \ref{sec6}.

The energy-momentum tensor of the boundary theory can be obtained from (\ref{graltmunu}), and turns out to be $\left\langle T_{\mu\nu}(x)\right\rangle=0$ $\forall$ odd $d$,
\begin{eqnarray}\label{tmunueven}
\left\langle T_{\mu\nu}\right\rangle&=&-\frac{L}{2\kappa^2}H^2{\rm diag}\,\left\{-(1-H^2r^2),\frac{1}{1-H^2r^2}\right\}\quad\text{for }d=2,\\
\left\langle T_{\mu\nu}\right\rangle&=&-\frac{L^3}{2\kappa^2}\frac{3H^4}{4} {\rm diag}\,\left\{ -(1-H^2r^2),\frac{1}{1-H^2r^2} , r^2,r^2 \sin^2 \theta     \right\}\quad\text{for }d=4.
\end{eqnarray}
This result captures the correct conformal anomaly present in even dimensions.

\subsection{The conformally flat patch}

To obtain a gravity solution dual to a QFT living in another dS chart we can proceed as we did in the previous section. However, note that for our previous background, the Fefferman-Graham metric (\ref{feffermangraham}) factorizes such that $g_{\mu\nu}(z,x)=f(z)g^{\text{dS}}_{\mu\nu}(x)$, with
\be\label{fdez}
f(z)=\left(1-\frac{H^2z^2}{4}\right)^2,
\ee
and $g^{\text{dS}}_{\mu\nu}(x)$ being the $d$-dimensional de Sitter metric. This fact provide us with a convenient shortcut: the bulk metric for any other dS chart can be obtained just by a coordinate transformation of (\ref{metric1}). We can thus express $g^{\text{dS}}_{\mu\nu}(x)$ in \emph{any} coordinate system and it will immediately related to our previous solution via a trivial diffeomorphism that does not mix the $z$-coordinate (and thus preserving the Fefferman-Graham form).

Here we are interested in the planar patch of de Sitter. The planar patch is conformally related to Minkowski space and it covers one half of global de Sitter,
\be
ds^2=\frac{1}{H^2\eta^2}\left(-d\eta^2+d\vec{x}^2\right).
\ee
Here $\eta$ is the so-called conformal time and is related to the usual FRW time by the relation
\be
d\eta\equiv\frac{dt}{a(t)}=\frac{dt}{e^{Ht}}.
\ee

For this coordinate system then, the corresponding bulk metric reads
\be
ds^2={L^2 \over z^2}\left(\frac{f(z)}{H^2\eta^2}\left(-d\eta^2+d\vec{x}^2\right)+dz^2\right),\label{metricMink}
\ee
where the function $f(z)$ is given in (\ref{fdez}).

The analysis of curvature invariants is the same as in equations (\ref{R1})-(\ref{R3}). The boundary stress-energy tensor is related to the one in the static patch by a trivial diffeomorphism:
\begin{eqnarray}
\left\langle T_{\mu\nu}\right\rangle&=&-\frac{L}{2\kappa^2 \eta^2}{\rm diag}\,\left\{-1,1\right\}\qquad\text{for }d=2,\\
\left\langle T_{\mu\nu}\right\rangle&=&-\frac{L^3}{2\kappa^2}\frac{3H^2}{4\eta^2} {\rm diag}\,\left\{ -1,1 , 1,1    \right\}\qquad\text{for }d=4
\end{eqnarray}
and $\left\langle T_{\mu\nu}(x)\right\rangle=0$ $\forall$ odd $d$.

\section{Entanglement entropy, mutual information and number of degrees of freedom\label{sec3}}
\begin{figure}[htbp]
\begin{center}
 \fbox{\includegraphics[height=5cm]{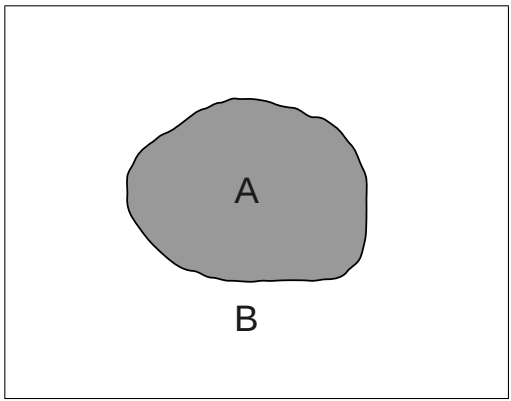}}
\caption{The total system can be divided into two subsystems A and $B\equiv A^c$; the entanglement entropy measures the amount of information loss because of smearing out in region B.}
\label{ss}
\end{center}
\end{figure}
When we consider an arbitrary quantum field theory with many degrees of freedom, we can ask about the entanglement of the system. To describe the system, we define a  density matrix, $\rho$, which is a self-adjoint, positive semi-definite, trace class operator.  Now, on a constant time Cauchy surface let us imagine dividing the system into two subsystems $A$ and $A^c$, where $A^c$ is the complement of $A$. The total Hilbert space then factorizes as $\cH_{\rm total} = \cH_{A} \otimes \cH_{A^c}$. For an observer who has access only to the subsystem $A$, the relevant quantity is the reduced density matrix defined as
\begin{eqnarray}
\rho_A = {\rm tr}_{A^c} \, \rho \ .
\end{eqnarray}

The entanglement between $A$ and $A^c$ is measured by the {\it entanglement entropy}, which is defined as the von Neuman entropy using this reduced density matrix
\begin{equation}
S_A = - {\rm tr}_A \, \rho_A \log \rho_A \ .
\end{equation}
\begin{figure}[htbp]
\begin{center}
 \includegraphics[angle=0, trim = 20mm 45mm 15mm 20mm, clip,width=0.60\textwidth]{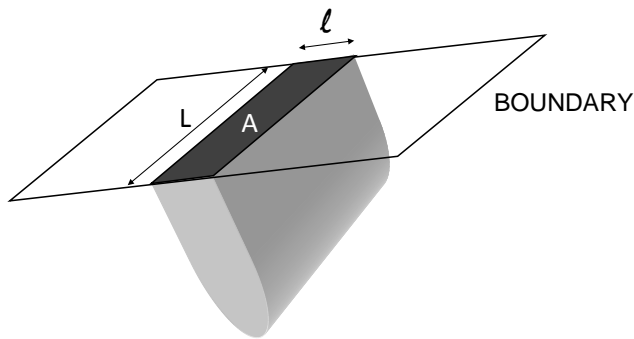}
\caption{A schematic diagram of the extremal surface used for calculation of the entanglement entropy.}
\label{hg}
\end{center}
\end{figure}
In AdS/CFT, Ryu and Takayanagi \cite{Ryu:2006bv} conjectured a formula to compute the entanglement entropy of quantum systems with a static $(d+1)$ dimensional gravitational dual. The entanglement entropy of a region $A$  of the quantum field theory is given by
\be \label{rt}
S_A = \frac{1}{4 G_N^{(d+1)}} {\rm min} \left[ {\rm Area} \left(\gamma_A \right)\right] \ ,
\ee
where $G_N$ is the bulk Newton's constant, and $\gamma_A$ is the $(d-1)$-dimensional area surface such that  $\partial \gamma_A = \partial A$. For background with time dependence, this proposal has been successfully generalized to \cite{Hubeny:2007xt}
\be \label{hrt}
S_A = \frac{1}{4 G_N^{(d+1)}} {\rm ext} \left[ {\rm Area} \left(\gamma_A \right)\right] \ ,
\ee
where now minimal area surface is replaced by extremal surface. It is well known that entanglement entropy of a spatial region in a local field theory is UV-divergent
\begin{equation}
S=S_{div}+S_{finite} \ .
\end{equation}
Only local physics contributes to the UV-divergent piece $S_{div}$.  On the other hand $S_{finite}$ contains information about the long range entanglement.

Mutual information is a quantity that is derived from entanglement entropy.
\begin{figure}[!]
\centering
\includegraphics[width=0.6\textwidth]{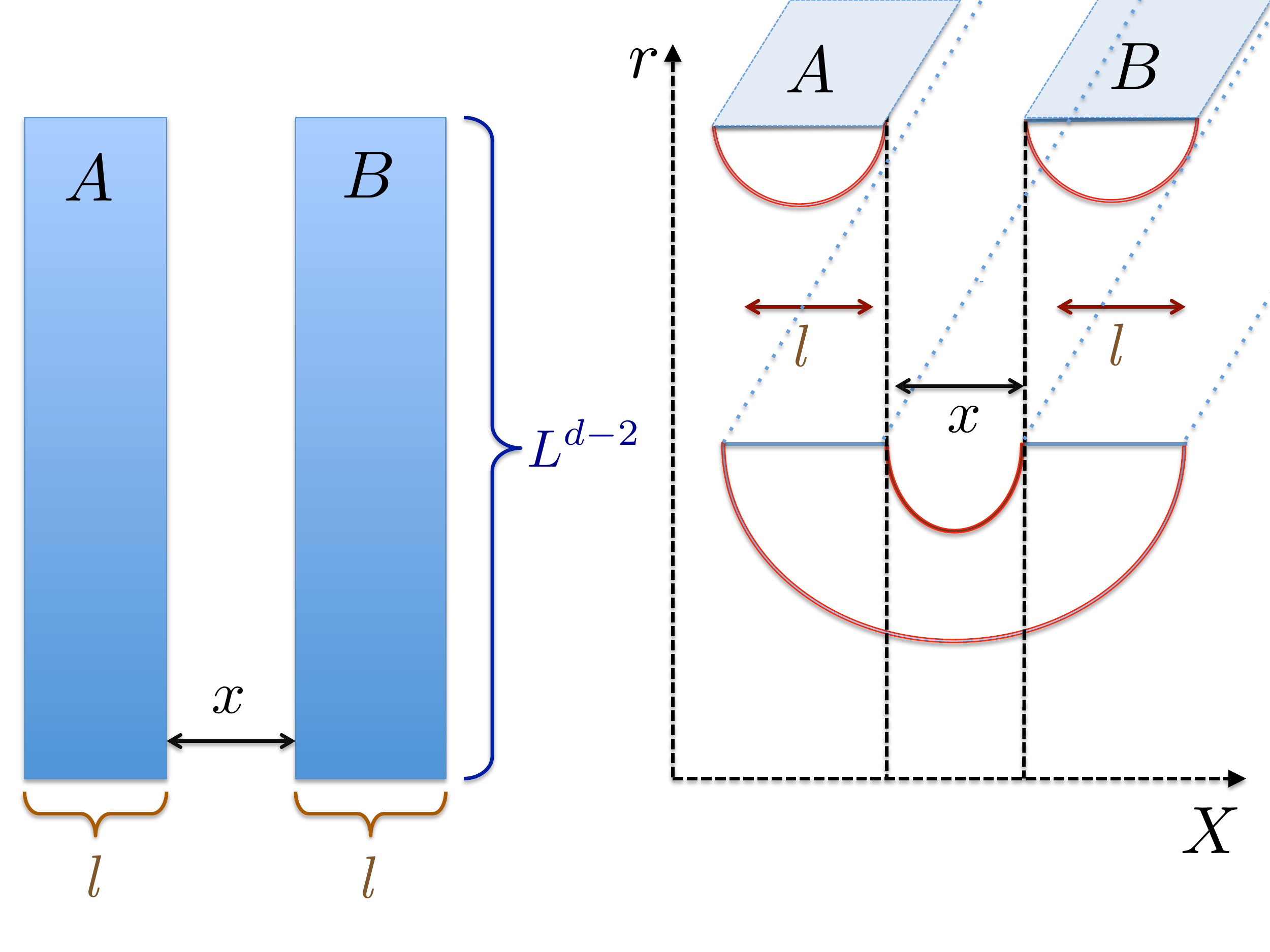}
\caption{The two disjoint sub-systems $A$ and $B$, each of length $l$ along $X$-direction and separated by a distance $x$. The schematic diagram on the right shows the possible candidates for minimal area surfaces which is relevant for computing $S_{A\cup B}$. See \cite{Fischler:2012uv} for a detailed discussion.}
\label{shape}
\end{figure}
Mutual information between two disjoint sub-systems A and B is defined as
\be
I(A,B)=S_A+S_B- S_{A\cup B}
\ee
where $S_A$, $S_B$ and $S_{A\cup B}$ denote entanglement entropy of the region $A$, $B$ and $A\cup B$ respectively with the rest of the system (see fig.~\ref{shape} for an example). Mutual information is a UV-finite quantity and hence it does not depend on regularization scheme. Moreover, as showed in \cite{PhysRevLett.100.070502}, given an operator $\O_A$ in the region $A$ and $\O_B$ in the region $B$, mutual information sets an upper bound
\begin{eqnarray} \label{mi1}
I(A, B) \ge \frac{\left(\langle \O_A \O_B \rangle - \langle \O_A \rangle \langle \O_B \rangle \right)^2}{2 \langle \O_A^2 \rangle \langle \O_B^2 \rangle}
\end{eqnarray}
and thus measures the total correlation between the two sub-systems: including both classical and quantum correlations.

Renormalized entanglement entropy, introduced in \cite{Liu:2012eea}, is another derived quantity which is UV-finite. For a region $A$ with a single length scale $R$, it is defined as
\begin{align}
\S_d(R)=& \frac{1}{(d-2)!!}\left(R\frac{d}{dR}-1\right)\left(R\frac{d}{dR}-3\right)...\left(R\frac{d}{dR}-(d-2)\right)S_d(R) \qquad  d ~\text{odd} \nonumber \\
=& \frac{1}{(d-2)!!}R\frac{d}{dR}\left(R\frac{d}{dR}-2\right)...\left(R\frac{d}{dR}-(d-2)\right)S_d(R) \qquad  d ~\text{even} \ ,
\end{align}
where $S_d(R)$ is the entanglement entropy of the region $A$. Renormalized entanglement entropy $\S_d(R)$ has some nice properties: (i) for a CFT $\S_d(R)$ is a constant, (ii) for a renormalizable quantum field theory both $\S_d(R\rightarrow 0)$ and $\S_d(R\rightarrow \infty)$ are constants and as $R$ is increased from zero to infinity $\S_d(R)$ interpolates between them, (iii) $\S_d(R)$ is expected to be most sensitive to degrees of freedom at scale $R$. $\S_d(R)$ somewhat naturally describes the RG flow of the entanglement entropy with distance scale \cite{Liu:2012eea, Liu:2013una}.

Particularly, for $d=2,3$ and $4$, $\S_d(R)$ is given by
\begin{align}
\S_2(R)=& R\frac{dS_2(R)}{dR}\ , \\
\S_3(R)=& R\frac{dS_3(R)}{dR}-S_3(R) \ , \\
\S_4(R)=& \frac{1}{2}\left(R^2\frac{d^2S_4(R)}{dR^2}-R\frac{dS_4(R)}{dR}\right)\ .
\end{align}

\section{Entanglement entropy in (1+1) dimensions\label{sec4}}
In this section we will compute the entanglement entropy of an interval $x\in \left[-\frac{a}{2},\frac{a}{2}\right]$ at time $\eta=\eta_0$ for a strongly coupled QFT living in the conformally flat patch of de Sitter in $(1+1)$ dimensions, using the holographic prescription (\ref{hrt}). We choose the standard Bunch-Davies vacuum state of the field theory and the dual bulk metric is given by (\ref{metricMink}). The minimal area surface can be parametrized by two functions: $x(z)$ and $\eta(z)$ with the boundary conditions
\begin{align}
 x(z=0)=\pm \frac{a}{2}, \qquad \eta(z=0)=\eta_0.
\end{align}
The area functional is given by,
\begin{equation}\label{action1}
\A=L \int  \frac{dz}{z} \sqrt{1+ \frac{f(z)}{H^2 \eta^2}\left(x'^2-\eta'^2 \right)}\ .
\end{equation}
Before, we proceed, let us perform a change of variable that will make our life easier
\begin{align}
\eta(z)=& -\frac{1}{H}\frac{e^{-H \tau(z)}}{\sqrt{1-H^2 r(z)^2}},\\
x(z)=& - H \eta(z) r(z).
\end{align}
Note that $\Delta r(0)=\Delta x(0)/ H |\eta_0|$ is the proper length of the interval $\Delta x(0)$. In terms of these new functions, the action becomes
\begin{equation}\label{action3}
\A=L \int  \frac{dz}{z} \sqrt{1+ f(z)\left(\frac{r'^2}{1-H^2 r^2}-(1-H^2 r^2)\tau'^2 \right)}
\end{equation}
with boundary conditions:
\begin{align}
 r(z=0)=\pm \frac{l}{2}, \qquad \tau(z=0)=\tau_0
\end{align}
where $\tau_0$ is related to $\eta_0$ and $l$ is the proper length of the interval $a$. The action is independent of $\eta$, however now we are restricted to $lH<2$, i.e. the interval is smaller than the size of the horizon.

Equations (\ref{action1}, \ref{action3}) lead to two conserved quantities:
\begin{align}
& \frac{f(z)(1-H^2 r^2)\tau'}{z \L}=c_1\ , \\
& \frac{f(z) x'}{z H^2 \eta^2 \L}=c_2\ ,
\end{align}
where, $c_1, c_2$ are constants and
\begin{equation}
\L=\sqrt{1+ \frac{f(z)}{H^2 \eta^2}\left(x'^2-\eta'^2 \right)}=\sqrt{1+ f(z)\left(\frac{r'^2}{1-H^2 r^2}-(1-H^2 r^2)\tau'^2 \right)}.
\end{equation}
Therefore, for $lH<2$, we obtain
\begin{equation}
\frac{d\tau}{dx}=e^{H \tau}\frac{c_1}{c_2}.
\end{equation}

\subsection{Connected solution}
When $lH<2$, we can use action (\ref{action3}) to obtain U-shaped solutions. From the action (\ref{action3}) it is clear that for these solutions $\tau(z)=\tau_0$ and only $r(z)$ has a nontrivial profile. Let us introduce $rH= \sin{\theta}$ and in terms of $\theta(z)$, equation of motion is obtained to be
\begin{equation}
H \theta'(z)=\pm \frac{H z \sqrt{f(z_c)}}{z_c\sqrt{f(z)^2-\frac{f(z_c)f(z)z^2}{z_c^2}}}
\end{equation}
and the area is given by
\begin{equation}
\A=2L\int_{\epsilon}^{z_c}\frac{dz}{z} \frac{f(z)}{\sqrt{f(z)^2-\frac{f(z_c)f(z)z^2}{z_c^2}}}=-2L \ln \epsilon+ 2L \ln \left(\frac{8 z_c}{4+H^2 z_c^2}\right),
\end{equation}
where, $z_c$ is the closest approach point obtained from the boundary condition
\begin{equation}
l=\frac{2}{H}\sin\left[\int_{0}^{z_c} \frac{dz H z \sqrt{f(z_c)}}{z_c\sqrt{f(z)^2-\frac{f(z_c)f(z)z^2}{z_c^2}}}\right] = \frac{8 z_c}{4+H^2 z_c^2}.
\end{equation}
Therefore, the entanglement entropy is given by,
\begin{equation}
S=\frac{c}{3} \ln \left(\frac{l}{\epsilon}\right)=\frac{c}{3} \ln \left(\frac{a}{\epsilon}\right)-\frac{c}{3}\ln(H |\eta_0|),
\end{equation}
where, $c=\frac{3L}{2 G_N^{(2+1)}}$.

\subsection{Disconnected solution}
\begin{figure}[!]
\centering
\includegraphics[width=0.80\textwidth]{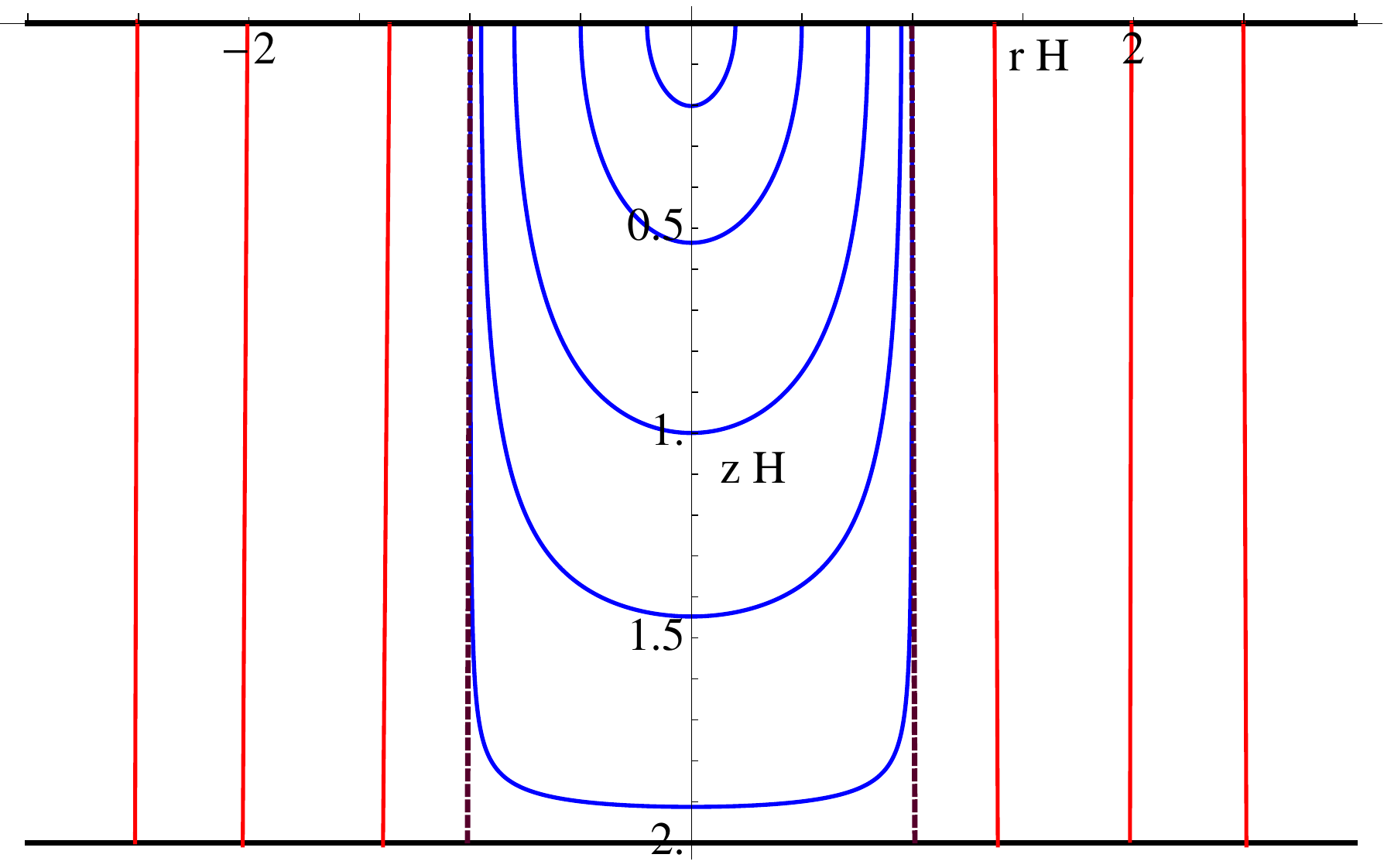}
\caption{Extremal surfaces for $d=2$ in $(r,z)$ coordinates; $zH=0$ is the boundary and $zH=2$ is the Killing horizon. Blue lines are U-shaped extremal surfaces for $lH<2$ and red lines are disconnected extremal surfaces for $lH>2$. Dashed brown line is the extremal surface for $lH = 2$.}
\label{extsur}
\end{figure}
The disconnected solution is given by : $x(z)=\pm \frac{a}{2}, \eta(z)=\eta_0$ and the corresponding entanglement entropy is given by
\begin{equation}
S=-\frac{c}{3}\log(\epsilon H)+\frac{c}{3} \log 2.
\end{equation}
Therefore, there is a transition from the connected solution to the disconnected solution at $lH=2$. It is important to note that $l<2/H$ is the region accessible to a single observer and hence only a {\it ``super-observer"} can ``see" this transition of entanglement entropy.

Finally the entanglement entropy is given by,
\begin{equation}
S=\frac{c}{3} \ln \left(\frac{l}{\epsilon}\right) \Theta(2-lH)+\frac{c}{3} \ln \left(\frac{2}{\epsilon H}\right) \Theta(lH-2),
\end{equation}
where, $\Theta$ is the Heaviside theta function. Few comments are in order: note that in $(1+1)$ dimensions, entanglement entropy of an interval of length $l$ in flat space is the same as entanglement entropy of an interval of proper length $l$ in the conformally flat patch of de Sitter, when the interval is smaller than the size of the horizon for a single observer. However, a ``super-observer" will see a difference in the behavior of the entanglement entropy.

\subsection{Mutual information}
\begin{figure}[htbp]
\begin{center}
 \includegraphics[height=8cm]{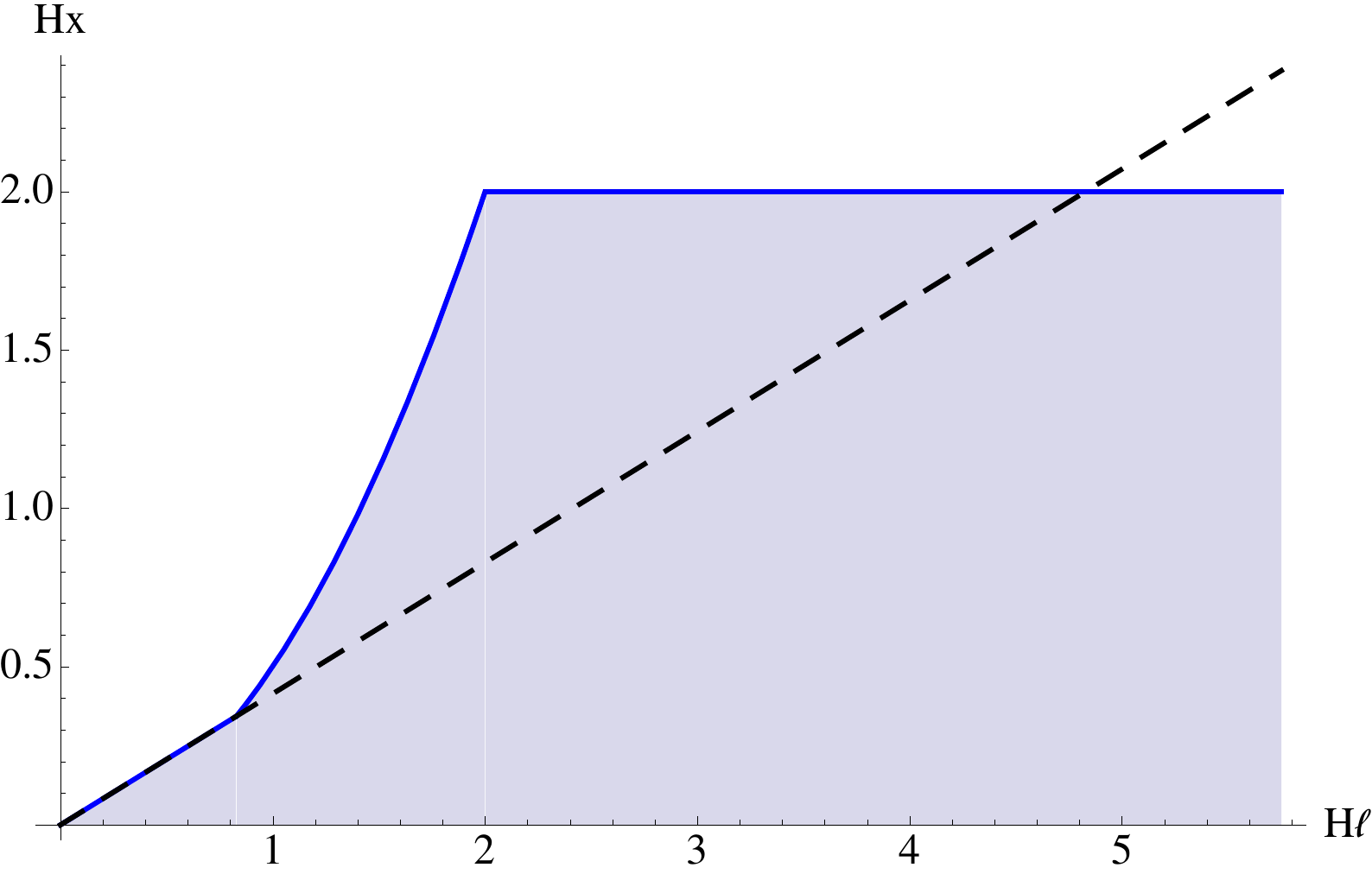}
\caption{Entanglement/disentanglement transition of mutual information in $(1+1)-$dimensions. Mutual information is nonzero only in the shaded region. Dashed black line is the transition curve in flat space.}
\label{2dmi}
\end{center}
\end{figure}

Let us now calculate mutual information of two intervals $A$ and $B$ of length $a$ (proper length $l$) separated by a distance $b$ (proper length $x$). Mutual information undergoes an interesting {\it entanglement/disentanglement ``phase-transition"} (see figure \ref{2dmi}) which is qualitatively different from  what has been discussed in \cite{Fischler:2012uv}. We will consider three separate cases:\\

\subsubsection{Case I: $x+2l\le 2/H$}
In this case, mutual information is independent of $H$:
\begin{align}
I(A,B)=&\frac{c}{3}\ln \left(\frac{l^2}{x(2l+x)}\right)=\frac{c}{3}\ln \left(\frac{a^2}{b(2a+b)}\right) \qquad \frac{b}{a}=\frac{x}{l}\le 0.414 \nonumber\\
=& 0 \qquad \frac{b}{a}=\frac{x}{l}> 0.414.
\end{align}
\subsubsection{Case II: $l\le 2/H$ and $x+2l\ge 2/H$}
\begin{align}
I(A,B)=&\frac{c}{3}\ln \left(\frac{H l^2}{2x}\right) \qquad Hl^2\ge 2x \nonumber\\
=& 0 \qquad Hl^2< 2x.
\end{align}
\subsubsection{Case III: $l\ge 2/H$}
\begin{align}
I(A,B)=&\frac{c}{3}\ln \left(\frac{2}{xH}\right) \qquad Hx\le 2 \nonumber\\
=& 0 \qquad Hx>2.
\end{align}

Let us now comment on a few key features of mutual information in this case. When the intervals are comparable to the horizon size (i.e. $l H\sim \O(1)$), entanglement between them increases. However, mutual information between any two intervals separated by a proper distance $xH\ge 2$ is identically zero. It implies that for any two intervals $A$ and $B$, separated by a distance larger than the horizon size, $\rho_{A\cup B}=\rho_A \otimes\rho_B$  and hence they are completely disentangled.

\subsection{Renormalized entanglement entropy}
For $d=2$, $\S_2(R)$ is monotonically decreasing for all Lorentz-invariant, unitary QFTs and hence it indeed describes RG flow of the entanglement entropy. It is interesting to investigate how $\S_2(R)$ behaves for de Sitter QFTs in the Bunch-Davies vacuum state. Particularly,  when $lH<2$, renormalized entanglement entropy is non-zero and independent of $lH$
\begin{equation}
\S_2(l)=\frac{c}{3}\ .
\end{equation}
However, for $lH>2$ it vanishes
\begin{equation}
\S_2(l)=0.
\end{equation}
Thus, the renormalized entanglement entropy undergoes a {\it phase transition}. This is not very surprising since $\S_2(l)$ measures entanglement correlations of a system at distance scale $l$ and two points separated by a distance $l>2/H$ are causally disconnected.

\section{Entanglement entropy of a sphere in $d-$dimensions\label{sec5}}
In this section, we will compute the entanglement entropy of a spherical region $\sum_{i=1}^{d-1}x_i^2 \le a^2$ for a field theory living in the conformally flat patch of $d-$dimensional de Sitter spacetime. We choose the Bunch-Davies vacuum state and the corresponding dual $(d+1)-$dimensional bulk metric is given by (\ref{metricMink}). We will show that the entanglement entropy of a sphere of radius $R$ in the static patch of de Sitter is the same as entanglement entropy of a sphere of proper radius $R$ in the conformally flat patch of de Sitter when the size of the sphere is smaller than the size of the horizon.

Because of the spherical symmetry, the extremal surface can be parametrized by two functions: $\rho(z)$ and $\eta(z)$ and the area functional is given by
\begin{equation}
\A=L^{d-1} \Omega_{d-1} \int  \frac{dz}{z^{d-1}} \frac{f(z)^{(d-2)/2}\rho^{d-2}}{(H\eta)^{d-2}}\sqrt{1+ \frac{f(z)}{H^2 \eta^2}\left(\rho'^2-\eta'^2 \right)}.
\end{equation}
where, $\Omega_{d-1}$ is the area of a unit sphere in $d-1$ dimensions. The boundary conditions are
\begin{align}
\eta(\epsilon)=\eta_0, \qquad \rho(\epsilon)=a\ ,
\end{align}
where, $\epsilon$ is the short distance cut-off that we need to introduce in order to regularize the area of the extremal surface. The proper radius of the sphere is given by, $R=a/(|\eta|H)$.

\subsection{$RH<1$: Connected solution}
When $\rho(z)/|\eta(z)|<1$, we can again perform a change of variables:\footnote{This change of variable is useful only for spherical regions.}
\begin{align}
\eta(z)=& -\frac{1}{H}\frac{e^{-H \tau(z)}}{\sqrt{1-H^2 r(z)^2}},\label{taueta}\\
\rho(z)=& - H \eta(z) r(z)\ ,
\end{align}
 where, $H r(z)<1$. In terms of these new functions, the action becomes
\begin{equation}\label{action2}
\A=L^{d-1} \Omega_{d-1}\int  \frac{dz}{z^{d-1}}f(z)^{(d-2)/2}r^{d-2} \sqrt{1+ f(z)\left(\frac{r'^2}{1-H^2 r^2}-(1-H^2 r^2)\tau'^2 \right)}
\end{equation}
with boundary conditions:
\begin{align}
 r(z=\epsilon)=R, \qquad \tau(z=\epsilon)=\tau_0
\end{align}
where $\tau_0$ is related to $\eta_0$ by equation (\ref{taueta}) and $R$ is the proper radius. Before proceeding further, a few comments are in order: First, note that the action (\ref{action2}) is independent of $\tau$ and hence it is clear that $\tau(z)=\tau_0$ is a solution. Also note that the action (\ref{action2}) is identical to that for a spherical region in the static patch of de Sitter, where the bulk metric is given by (\ref{metric1}). Therefore, entanglement entropy of a sphere of radius $R$ in the static patch of de Sitter with the observer at the center of the sphere, is the same as entanglement entropy of a sphere of proper radius $R$ in the conformally flat patch of de Sitter, provided the size of the sphere is smaller than the size of the horizon. However, it is not very clear how to generalize this result for non-spherical regions because boundary conditions are different in different coordinates.  In the conformally flat patch, temporal boundary condition is given by $\eta=$constant, whereas, in the static patch it is given by $\tau=$constant. These two boundary conditions are identical only for spherical regions.

Now we will  obtain the equation of motion for the radial profile using the action (\ref{action2})
\begin{align}
&r^3 \left[r'\left(2 (d-1) f-d z f'\right)-2 z f r''\right]+2 z r^2 \left[(d-1) f r'^2+2 (d-2)\right]-2 (d-2) z \left[f r'^2+1\right]\nonumber\\
&+r \left[r' \left(-(d-1) f \left(2 f-z f'\right) r'^2+d z f'-2 d f+2 f\right)+2 z f r''\right]-2 (d-2) z r^4=0\ .
\end{align}
It has a simple solution:
\begin{equation}
r(z)=\sqrt{R_0^2-\frac{z^2(1-H^2 R_0^2)}{f(z)}}\ ,
\end{equation}
where $R_0$ is a constant which can be computed using the boundary condition  $r(\epsilon)=R$, yielding
\begin{equation}
R=\sqrt{R_0^2-\frac{\epsilon^2(1-H^2 R_0^2)}{f(\epsilon)}}.
\end{equation}
Now the area is given by,
\begin{align}\label{area}
\A=\frac{L^{d-1} \Omega_{d-1} R_0}{z_c} \int_{\epsilon/z_c}^{1} \frac{du}{u^{d-1}} \sqrt{f(u)}\left(R_0^2 \frac{f(u)}{z_c}-u^2\right)^{(d-3)/2},
\end{align}
where,
\begin{equation}
z_c=\frac{2 \left(1-\sqrt{1-H^2 R_0^2}\right)}{H^2 R_0}, \qquad f(u)= \left(1-\frac{H^2 z_c^2 u^2}{4}\right)^2,
\end{equation}
such that $r(z_c)=0$.

\subsection{$RH>1$: Disconnected solution}
\begin{figure}[!]
\centering
\includegraphics[width=0.95\textwidth]{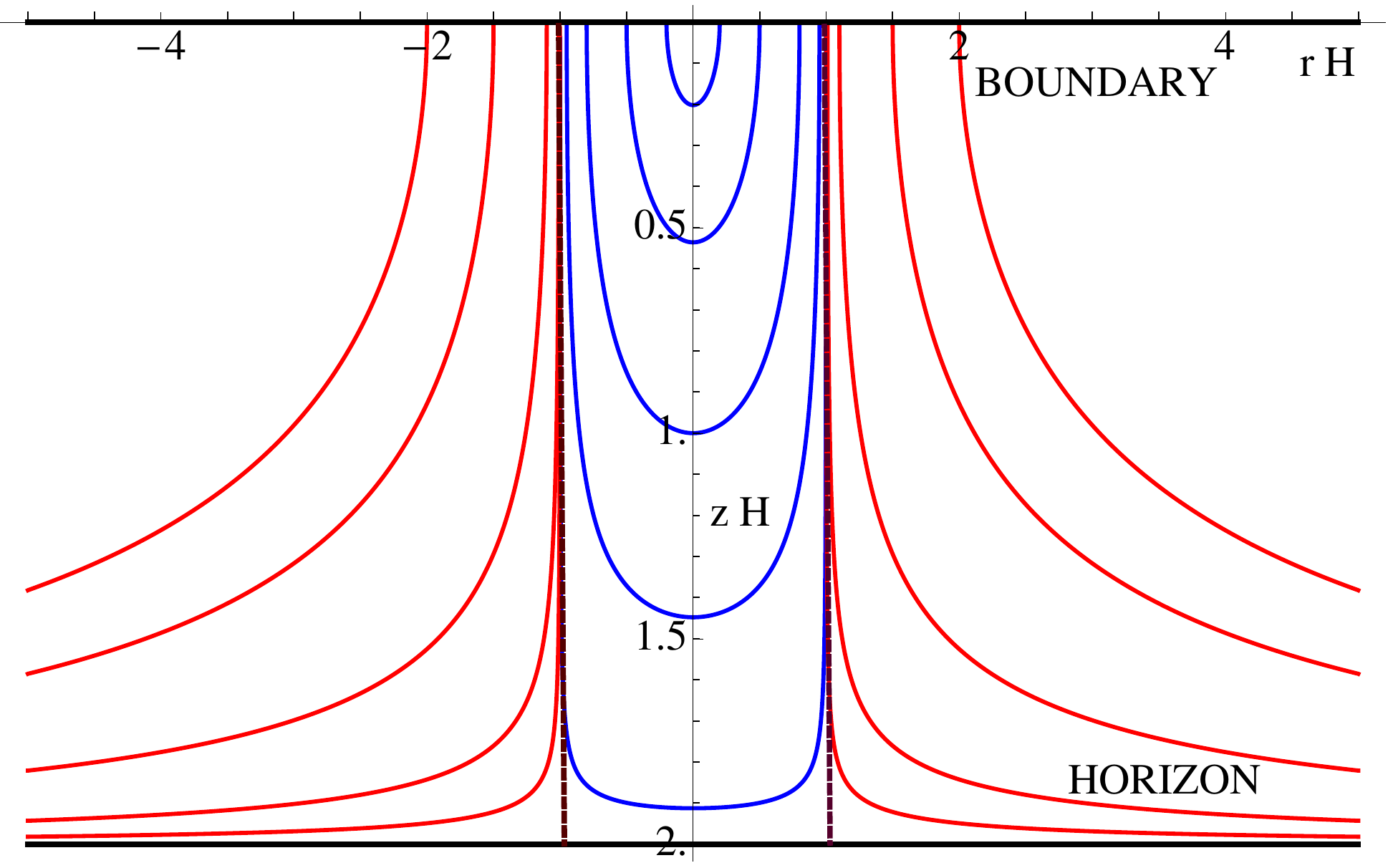}
\caption{Extremal surfaces for $d>2$ in $(r,z)$ coordinates; $zH=0$ is the boundary and $zH=2$ is the Killing horizon. Blue lines are extremal surfaces for $RH<1$ and red lines are extremal surfaces for $RH>1$. Dashed brown line is the extremal surface for $RH=1$. Note that for $RH>1$, extremal surfaces do not penetrate the region inside $RH=1$.}
\label{extsur}
\end{figure}
Now we will do the following change of variable,
\begin{align}
\eta(z)=& -\frac{1}{H}\frac{e^{-H \tau(z)}}{\sqrt{H^2 r(z)^2-1}},\\
\rho(z)=& - H \eta(z) r(z).
\end{align}
 This change of variable is valid only for $H r>1$. In terms of these new functions, the action becomes
\begin{equation}
\A=L^{d-1} \Omega\int  \frac{dz}{z^{d-1}}f(z)^{(d-2)/2}r^{d-2} \sqrt{1+ f(z)\left(-\frac{r'^2}{H^2 r^2-1}+(H^2 r^2-1)\tau'^2 \right)}.
\end{equation}
The boundary conditions are
\begin{align}
 r(z=0)=R, \qquad \tau(z=0)=\tau_0
\end{align}
with $RH>1$ and $\tau_0$ is related to $\eta_0$ and $R$ is the proper radius. Note that the action is again independent of $\tau$.  Solutions of the equations of motion are
\begin{align}
\tau(z)=&\tau_0,\\
r(z)=&\sqrt{R_0^2+\frac{z^2(H^2 R_0^2-1)}{f(z)}}.
\end{align}
Because the area of the extremal surface is divergent, we will again introduce a short distance cut-off $z=\epsilon$ such that $r(\epsilon)=R$. Therefore,
\begin{equation}
R=\sqrt{R_0^2+\frac{\epsilon^2(H^2 R_0^2-1)}{f(\epsilon)}}.
\end{equation}
Behaviors of the extremal surfaces for $RH<1$ and $RH>1$ are entirely different and for $RH>1$, extremal surfaces do not penetrate the region inside $RH=1$; this feature is schematically shown in figure \ref{extsur}.\footnote{A similar behavior of the extremal surfaces has also been observed for Kasner-AdS soliton background \cite{Engelhardt:2013jda}.} Also note that the behavior of the extremal surfaces near the Killing horizon is rather unique, since the near horizon region of the bulk contributes insignificantly to the area of the surfaces.\footnote{For a discussion on behavior of the extremal surfaces near an event horizon see \cite{Hubeny:2012ry,Fischler:2012ca}.} Presence of the Killing horizon is the bulk reflection of the boundary causal structure and it is responsible for the  phase transition of the entanglement entropy.

\subsection{Entanglement entropy in $(2+1)$-dimensions}
For $H=0$ and $d=3$, we recover the flat space results:
\begin{equation}
\A=2 \pi  L^{2}  R \left(\frac{1}{\epsilon}-\frac{1}{R}\right).
\end{equation}
For non-zero $H$ and $RH<1$, we obtain
\begin{equation}
\A=2 \pi  L^{2}  R \left(\frac{1}{\epsilon}-\frac{1}{R}\right).
\end{equation}
Therefore, for $RH<1$
\begin{equation}
S_H(R)=S_0(R)=2 \pi  c_3  R \left(\frac{1}{\epsilon}-\frac{1}{R}\right),
\end{equation}
where, $c_3=L^2/4G^{3+1}_N$. The fact that $S_H(R)=S_0(R)$, for regions smaller than the size of horizon, is probably related to the absence of conformal anomaly in odd dimensions.

For $RH>1$, entanglement entropy is an area-worth quantity
\begin{equation}
S_H(R)=2 \pi  c_3  R \left(\frac{1}{\epsilon}-H\right),
\end{equation}
There is a ``phase transition" of the entanglement entropy at $RH=1$ where the first derivative of the entanglement entropy is discontinuous:
\begin{equation}
S_H '(RH\rightarrow 1)|_- - S_H '(RH\rightarrow 1)|_+ =2 \pi  c_3 H\ .
\end{equation}

\subsection{Entanglement entropy in $(3+1)$-dimensions}
For $H=0$ in $d=4$, we have
\begin{equation}
\A= \pi  L^{3}   \left(\frac{2 R_0^2}{\epsilon^2}+2\log\left(\frac{\epsilon}{2R_0}\right)-1\right)=\pi  L^{3}   \left(\frac{2 R^2}{\epsilon^2}+2\log\left(\frac{\epsilon}{2R}\right)+1\right).
\end{equation}
For non-zero $H$ and $RH<1$, we obtain
\begin{align}
\A &= \pi  L^{3}   \left(\frac{2 R_0^2}{\epsilon^2}+2\log\left(\frac{\epsilon}{2R_0}\right)+R_0^2 H^2-1\right) \nonumber \\
&=\pi  L^{3}   \left(\frac{2 R^2}{\epsilon^2}+2\log\left(\frac{\epsilon}{2R}\right)-R^2 H^2+1\right).
\end{align}
Hence,
\begin{equation}
S_H(R)-S_0(R)=-c_4 ~\pi R^2 H^2,
\end{equation}
where, $c_4=L^3/4G^{4+1}_N$. Note that in $d=4$, energy momentum tensor of the boundary theory $\langle T_{\mu \nu}(x)\rangle \ne 0$. It is also interesting to note that at finite temperature in flat space-time $S_T(R)-S_0(R)\sim R^4 T^4$. Whereas in de Sitter at strong coupling $S_H(R)-S_0(R)\sim-R^2 T_{dS}^2$ and hence the behavior is completely different from what one would obtain for the same field theory in flat space-time at $T=T_{dS}$.

For $RH>1$, the entanglement entropy is given by
\begin{align}
S_H(R)= \pi  c_4   \left(\frac{2 R^2}{\epsilon^2}+2\log\left(\frac{\epsilon}{2R}\right)+1-H^2R^2-  2 RH \sqrt{-1+H^2R^2}\right. \nonumber\\
\left. +2\log\left[ RH+ \sqrt{H^2 R^2-1}\right]\right)\ .
\end{align}
Again there is a {\it  phase transition} at $RH=1$. This phase transition is a signal of a drastic change in correlations at distance $R=1/H$. It is important to note that $R < 1/H$ is the region accessible to a single observer and hence only a {\it Òsuper-observerÓ} can ``see" this phase transition of the entanglement entropy.

In the limit $R >>1/H$, we obtain
\begin{align}
S_H(R)\approx \pi  c_4   \left(\frac{2 R^2}{\epsilon^2}+2\log\left(\epsilon H\right)-3 H^2R^2\right)\ ,
\end{align}
and hence the finite piece is proportional to the proper area of the sphere. Note that in the limit $R >>1/H$, we do not have a term proportional to the number of e-foldings mainly because our extremal surfaces are disconnected in this limit. If one imposes smoothness of the solution as an additional criterion for it to be a good Ryu-Takayanagi surface, then in the limit $R >>1/H$  one will get an extra term which is proportional to the number of e-foldings \cite{Maldacena:2012xp}.

\subsection{Renormalized entanglement entropy}
The entanglement entropy of a sphere is continuous but not smooth at $R=1/H$. As a consequence, the renormalized entanglement entropy undergoes a {\it phase transition} at $R=1/H$ where it is discontinuous; however a single observer will never see this phase transition.
\subsubsection{$d=3$}
For the vacuum of Lorentz invariant, unitary QFTs $\S_3(R)$ is monotonically decreasing and non-negative \cite{Casini:2012ei}, providing a central function for the F-theorem. It is interesting to investigate how $\S_3(R)$ behaves for de Sitter QFTs in the Bunch-Davies vacuum state.  When $RH<1$, renormalized entanglement entropy is given by,
\begin{equation}
\S_3(R)=2\pi c_3.
\end{equation}
For $RH>1$,
\begin{equation}
\S_3(R)=0.
\end{equation}
Thus, the renormalized entanglement entropy undergoes a {\it phase transition}. This is expected since $\S_3(R)$ measures entanglement correlations of a system at distance scale $R$ and two regions separated by a distance $R>1/H$ are causally disconnected.

\subsubsection{$d=4$}
In $d\ge 4$ dimensions, the renormalized entanglement entropy is more complicated, since there are indications that it is neither monotonic nor non-negative \cite{Liu:2012eea}.\footnote{It is known that $\S_4(R)$ is non-monotonic and negative for several systems; for example GPPZ flow which describes the flow of $\N=4$ SYM to a confining theory under a mass deformation (which has UV-dimensions $\Delta=3$)\cite{Liu:2012eea}.}   However it is still expected that $\S_4(R\rightarrow 0)>\S_4(R\rightarrow \infty)$ because of the a-theorem \cite{Komargodski:2011vj,Komargodski:2011xv}. A similar behavior is observed for QFTs in de Sitter space-time. For $RH<1$, renormalized entanglement entropy is a constant
\begin{equation}
\S_4(R)=2\pi c_4.
\end{equation}
The renormalized entanglement entropy undergoes a {\it phase transition} at $R=1/H$ and for $RH>1$, $\S_4(R)$ is neither positive nor monotonically decreasing function of $R$
\begin{equation}
\S_4(R)=2 c_4 \pi  \left(1-\frac{RH}{\sqrt{R^2H^2-1}}\right).
\end{equation}
However, $\S_4(R\rightarrow 0)>\S_4(R\rightarrow \infty)$ still holds. One perhaps can argue that the behavior of $\S_4(R)$ is still acceptable because non-monotonic, non-positive regions are not accessible to a single observer. It is also possible that non-positive, non-monotonic behavior of $\S_4(R)$ is an indication that the definition of $\S_4(R)$ should be modified in order to construct a central function out of the entanglement entropy.

\section{Thermalization of dS QFTs\label{sec6}}

In this section we are interested in studying entanglement entropy in a thermalizing state of the same dS QFTs. We start by revisiting some generalities of the solutions found in \cite{Marolf:2010tg} for holographic theories of dS with $T\neq T_{\text{dS}}$ and then we construct the time-dependent generalization of these geometries. We can think of these bulk solutions as the Vaidya version of the so-called hyperbolic black holes described in \cite{Emparan:1998he,Birmingham:1998nr,Emparan:1999gf}.

\subsection{Hyperbolic black holes and dS QFTs with $T\neq T_{\text{dS}}$}

The basic idea goes as follows: the static patch of dS$_d$ is conformally related to the Lorentzian hyperbolic
cylinder $\mathbb{R}\times\mathbb{H}_{d-1}$ where $\mathbb{H}_{d-1}$ is the Euclidean hyperboloid,
\be
ds^2 = (1-H^2\,r^2) \, \left(-dt^2 + \frac{dr^2}{(1-H^2\, r^2)^2} + \frac{r^2}{1-H^2\,r^2}\, d\Omega_{d-2}^2\right)\,.
\label{confhc}
\ee
Moreover, holographic duals for theories living in the hyperbolic cylinder with Euclidean time period given by $\beta=1/T$ are know. These are the so-called hyperbolic (or topological) black holes,
\begin{equation}
ds^2 = -f(\rho) \, dt^2 + \frac{d\rho^2}{f(\rho)} + \rho^2 \, d\Sigma_{d-1}^2  \,,
\label{hypbh}
\end{equation}
\be
f(\rho ) = \frac{\rho^2}{L^2} - 1 - \frac{\mu}{\rho^{d-2}}\,,
\ee
where $d\Sigma_{d-1}^2 = d\xi^2 + \sinh^2 \xi\, d\Omega_{d-2}^2$ is the metric of the Euclidean hyperboloid. Alternatively, the mass can be given as
\be
\mu=\rho_+^{d-2}\left(  \frac{\rho_+^2}{L^2}-1\right)\,,
\ee
where $\rho_+$ is the radius of the event horizon, given by the
largest positive positive root of $f(\rho)$. To find the Hawking temperature $T$ for this solution, we must impose that the Euclidian time $t_{\text{E}}$ be periodically identified with appropriate period, $t_{\text{E}}\sim t_{\text{E}} + \beta$. A simple calculation shows that
\be\label{tempds}
T=\frac{1}{4\pi}\frac{d}{d\rho}f(\rho)\bigg|_{\rho_+}=\frac{\rho_+^2 d-(d-2)L^2}{4\pi L^2 \rho_+}\,.
\ee
This relation can be inverted to find
\be
\rho_+=\frac{2\pi T L^2}{d}\left[1+\sqrt{1+\frac{d(d-2)}{4\pi^2T^2L^2}}\right]\,,
\ee
which allows us to take $T$ as the parameter that determines the solution. The case $\mu=0$ is isometric to
AdS and is not properly a black hole as it is completely non-singular. However, it covers a smaller portion of the entire manifold and the coordinate patch breaks down at $\rho = \rho_+ = L$. The Killing horizon in this case is analogous to a Rindler horizon, with associated inverse
temperature $\beta = 2\pi L$, and non-vanishing area. Solutions with $\mu\neq0$ possess a true singularity at $\rho=0$.

\begin{figure}[!htm]
\begin{center}
\includegraphics[angle=0,
width=0.65\textwidth]{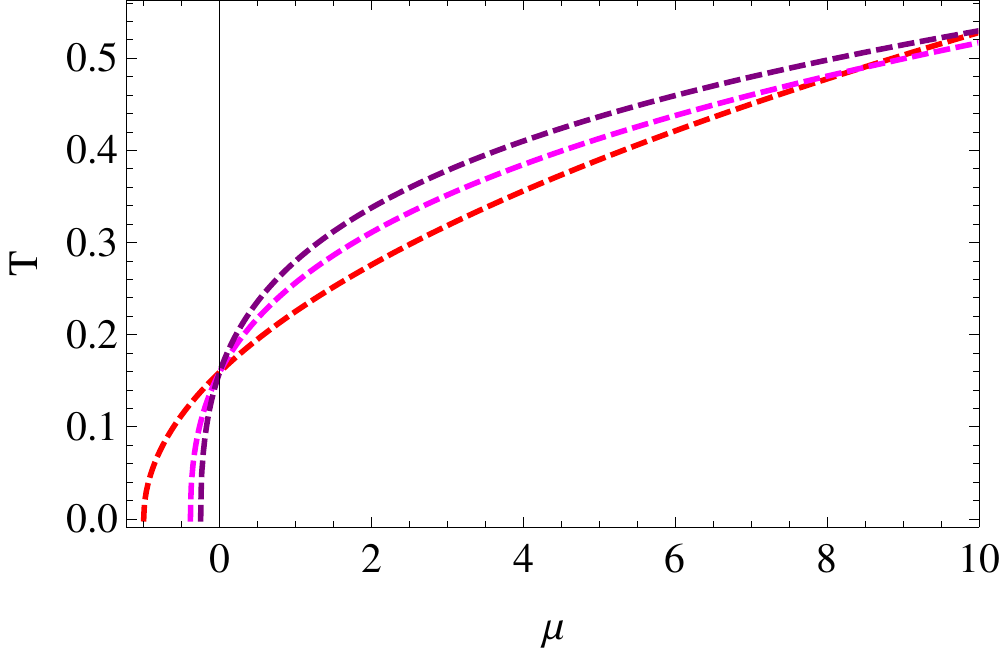}
\caption{\small Hawking temperature as a function of the black hole mass for $d=2$ (red), $3$ (pink) and $4$ (purple). In order to make the plots we have set $L=1$ so both $T$ and $\mu$ are measured in units of the AdS radius.}
\label{temp}
\end{center}
\end{figure}

In contrast with the regular AdS black holes, the zero temperature solution of (\ref{hypbh}) is different from the one that is isometric to AdS. In fact,
there is a range of negative values for $\mu$ such that the solutions still possess regular horizons and have sensible thermodynamics. The minimum values of $\mu$ and $\rho_+$ allowed, for which the horizon degenerates, are
\be
\mu^{\text{ext}}=-\frac{2}{d-2}\left(\frac{d-2}{d}\right)^{d/2}L^{d-2}\,,\qquad\rho_+^{\text{ext}}=\sqrt{\frac{d-2}{d}}L\,.
\ee
For these values the corresponding black hole is extremal. In general, the Hawking temperature is monotonically increasing with respect to the black hole mass. In Figure \ref{temp} we can see this behaviour for black holes of various dimensions. Interestingly, the Penrose diagram for a
hyperbolic black hole with negative $\mu$ is like that of a Reissner-Nordstr\"{o}m-AdS black hole. For
positive $\mu$ it is instead like that of a Schwarzschild-AdS black hole \cite{toporefs1,toporefs2,toporefs3,toporefs4,toporefs5,toporefs6,toporefs7}.

Now, the field theory dual to the hyperbolic black hole (\ref{hypbh}) lives in the hyperbolic cylinder which, according to (\ref{confhc}), is conformal to the static patch of de Sitter.  The fact that we are dealing with a CFT presents
us with an interesting possibility: among the infinite number of conformal frames available to us, we can choose the one related to $\mathbb{R}\times\mathbb{H}_{d-1}$ via the specific Weyl transformation $ds^2\to (1-H^2r^2)ds^2$. As first explained in \cite{Imbimbo:1999bj}, Weyl transformations in the boundary are dual to specific bulk diffeomorphism.\footnote{In fact, in the context of AdS/CFT a given bulk metric is understood to induce not a specific boundary metric, but a specific boundary conformal structure \cite{Witten:1998qj}.} In our case, the desired transformation can be achieved by defining a new coordinate
\begin{equation}
r = \frac{1}{H} \, \tanh\xi\,,
\label{}
\end{equation}
which brings the metric (\ref{hypbh}) into the form
\begin{equation}
ds^2 = \frac{H^2\, \rho^2}{1-H^2\,r^2} \, \left(-\frac{f(\rho)}{H^2\,\rho^2}\, (1 -H^2\,r^2)\, dt^2 + \frac{dr^2}{1-H^2\,r^2} + r^2\, d\Omega_{d-2}^2 \right) + \frac{d\rho^2}{f(\rho)}\,.
\label{dsdduals}
\end{equation}
This bulk geometry is now dual to a CFT on the static patch of dS$_d$ at a temperature given by (\ref{tempds}), which does not have to coincide with the de Sitter temperature $T_{\text{dS}}$. For the particular case $T=T_{\text{dS}}$, the background (\ref{dsdduals}) differs from (\ref{metric1}) only by a simple coordinate transformation.

From the boundary theory perspective, we are taking the period of the Euclidean time to be $1/T$ instead of $2\pi/H$. Of course,
unless $T=T_{\text{dS}}$, the Euclidean manifold
has a conical singularity which map in the Lorentzian signature to the de Sitter horizon.
Nevertheless, as long as we restrict our attention to study the physics inside the horizon, there is
nothing wrong about the corresponding Lorentzian spacetime. Physically, we can think of this as having
an external heat bath sitting at the de Sitter horizon which keeps
the system at a temperature different from $T_{\text{dS}}$. Therefore, the geometry does not
smoothly continue through the horizon because we run into the thermal bath, which is visible
as the conical singularity in the Euclidean space. The nature of this external bath becomes clear upon inspection of the holographic stress tensor, which implies that it can be identified as a radiation source. More specifically, separating the trace-free and trace parts, the energy-momentum tensor of the boundary theory is found to be \cite{Marolf:2010tg}:
\bea
T_\mu^{\, \nu} &=&\frac{L H^2}{2\kappa^2} \left[\frac{{\cal F}_2(T,H) }{1-H^2\,r^2} \;  \text{diag}\big\{-1,1 \big\} +  \text{diag} \big\{1,1 \big\} \right]\quad \text{for }d=2\,,\\
 T_\mu^{\, \nu} &=& \frac{L^{3}H^4}{2\kappa^2} \,\left[\frac{{\cal F}_4(T,H)}{(1-H^2\,r^2)^2}  \; \text{diag}\big\{-3,1,1,1\big\} +\frac{3}{4}\,\text{diag}\big\{1,1,1,1\big\}\right]\quad \text{for }d=4\,,
\eea
while
\begin{equation}
T_\mu^{\, \nu} =  \frac{L^{d-1}H^d}{2\kappa^2} \left[\frac{{\cal F}_d(T,H)}{(1-H^2\,r^2)^{\frac{d}{2}}}\;  \text{diag}\left\{1-d,1,1,\cdots,1\right\}\right]\quad\forall\text{ odd }d\,.
\label{Todddds}
\end{equation}
The function ${\cal F}_d(T,H)$ has the property that it vanishes for $T=T_{\text{dS}}$.

Thus, as we can see, the holographic stress tensor provides a smooth interpolation between the theory
on de Sitter space at temperature $T_{\text{dS}}$ and finite temperature physics in Minkowski space. The extra piece that appears for even dimensions comes naturally upon integration of the gravitational conformal anomaly.

\subsection{Gravitational collapse and black hole formation}

Our goal is to study the approach to thermalization of entanglement entropy in a time-dependent setup. The time-dependent configuration we are looking for here should capture the physics of gravitational collapse in the bulk which translates into the study of black hole formation. Thus, we will construct a generalized version of the Vaidya spacetimes for the so-called hyperbolic black holes.

To find the corresponding background, we have to couple the action in (\ref{action0}) with an external source
\be
S = S_0 + \alpha\, S_{\rm ext} \ ,
\ee
where $\alpha$ is a constant. For the physics we want to study in the present context we do not need to specify the form of $S_{\rm ext}$. The equations of motion in this case will take the following form
\be
R_{\mu\nu} - \frac{1}{2} \left(R- 2 \Lambda \right) g_{\mu\nu} = 2\alpha\kappa^2\, T_{\mu\nu}^{\rm ext}\,.
\ee
These field equations lead to the well-known AdS-Vaidya solution. For black holes with planar horizons, the relevant physics is restricted to the Poincar\'{e} patch and the metric takes the form
\be \label{vaid1}
ds^2 = \frac{L^2}{z^2} \left( - f(v,z) dv^2 - 2 dv dz + d\vec{x}^2 \right)\,,
\ee
\be
f(z,v) = 1 - m(v) z^d\,.
\ee
This background has been used extensively in the program of holographic thermalization \cite{AbajoArrastia:2010yt,Balasubramanian:2010ce,Balasubramanian:2011ur}. The metric (\ref{vaid1}) is written in terms of Eddington-Finkelstein coordinates (so that $v$ labels ingoing null trajectories) and represents a $(d+1)$-dimensional infalling shell of null dust.\footnote{For $m(v)=M=constant$, the metric (\ref{vaid1}) reduces to the usual AdS-Schwarzschild black hole.}
To see this directly, let us analize the matter contents that leads to the above solution. The function $m(v)$ is arbitrary and captures the information of the black hole formation. Quite generally, $m(v)$ is chosen to interpolate between zero as $v\to -\infty$ (corresponding to pure AdS) and a constant value as $v\to \infty$ (corresponding to AdS-Schwarzschild). A particular example of such a function is
\begin{eqnarray} \label{mchange}
m(v) = \frac{M}{2} \left(1 + \tanh \frac{v}{v_0} \right) \,,
\end{eqnarray}
where $v_0$ is a parameter that fixes the thickness of the shell. With this choice, the external source must yield the following energy-momentum tensor
\begin{eqnarray}
2\alpha\kappa^2\, T_{\mu\nu}^{\rm ext}  = \frac{(d-1)z^{d-1}}{2L^{2(d-1)}}  \frac{dm}{dv} \delta_{\mu v} \delta_{\nu v} \ .
\end{eqnarray}
If we identify $k_\mu = \delta_{\mu v}$, then we get \cite{AbajoArrastia:2010yt}
\begin{eqnarray}
T_{\mu\nu}^{\rm ext} \sim k_\mu k_\nu \ , \quad {\rm with} \quad k^2 = 0 \ ,
\end{eqnarray}
which is characteristic of null dust.

We can proceed in the same way and generalize the hyperbolic black holes (\ref{dsdduals}) to include a time-dependent mass. We find it convenient to define an inverse radius $z=L^2/\rho$, so that the boundary of AdS now lies at $z\to0$. Going to Eddington-Finkelstein coordinates, where the coordinate $v$ is defined as
\begin{eqnarray}
dv = dt - \frac{dz}{f(z,v)}\,,
\end{eqnarray}
we obtain\footnote{Here, we have rescaled the mass according to $m=\mu/L^{2(d-1)}$.}\footnote{From here on we will set the AdS radius $L$ equal to 1. This implies that all dimensionful quantities will be measured in units of this scale.}
\be
ds^2 = \frac{L^2}{z^2}\left(-f(z,v)dv^2-2dvdz+\frac{H^2\, L^2}{(1-H^2\,r^2)^2}dr^2  +\frac{H^2\, L^2 }{(1-H^2\,r^2)} r^2 d\Omega_{d-2}^2\right)\,,
\ee
\be
f(z,v) = 1 - m(v) z^d - \frac{z^2}{L^2}\,.
\ee
Similar to the planar AdS-Vaidya (\ref{vaid1}), the energy-momentum tensor of the external source yields
\begin{eqnarray}
2\alpha\kappa^2\, T_{\mu\nu}^{\rm ext}=\frac{d-1}{2}z^{d-1}  \frac{dm}{dv} \delta_{\mu v} \delta_{\nu v} \,,
\end{eqnarray}
which implies that the infalling shell is made of null dust.

In order to avoid possible issues in the boundary theory, we want to constrain the mass function in order to satisfy the null energy condition in the bulk.\footnote{The contrary is known to lead to violations of the strong subadditivity inequality \cite{Callan:2012ip,Wall:2012uf,Caceres:2013dma}.} This implies in particular that $m(v)$ should be an increasing function of $v$. In the context of the hyperbolic black holes we are considering here, however, there is a novel possibility that is not present in the traditional AdS-Vaidya backgrounds: we can start in a state with zero mass and end up with a finite positive mass or we can start with a negative mass (equal or above the extremal value) and finish with an increased mass (positive or negative). For concreteness, and with the aim of studying a situation of physical relevance, we will focus in the following mass function:
\be\label{m1fun}
m_1(v) = \frac{M_1}{2} \left(1 - \tanh \frac{v}{v_0} \right) \,,
\ee
where $M^{\text{ext}}\leq M_1<0$. In the boundary theory, this choice is equivalent to prepare the system in a state with $T<T_{\text{dS}}$ and then letting it evolve to the Bunch-Davies vacuum; in other words, the system undergoes a period of particle production and ends as a thermal bath of quanta at temperature $T=T_{\text{dS}}$. To be even more specific, we will set
\be\label{M1M2val}
M_1=M^{\text{ext}}=-\frac{2}{d-2}\left(\frac{d-2}{d}\right)^{d/2}\,,
\ee
so that the initial state is at zero temperature. We will see later that the relevant physics we are interested here is not affected by the specific value of $M_1$.

We can also consider a mass function of the type,
\be\label{m2fun}
m_2(v) = \frac{M_2}{2} \left(1 + \tanh \frac{v}{v_0} \right) \,,
\ee
with $M_2>0$. This corresponds to a situation in which we start in the Bunch-Davies vacuum and then evolve to a state with $T>T_{\text{dS}}$. We will briefly comment on this possibility.

\begin{figure}[!htm]
\begin{center}
\includegraphics[angle=0,
width=0.65\textwidth]{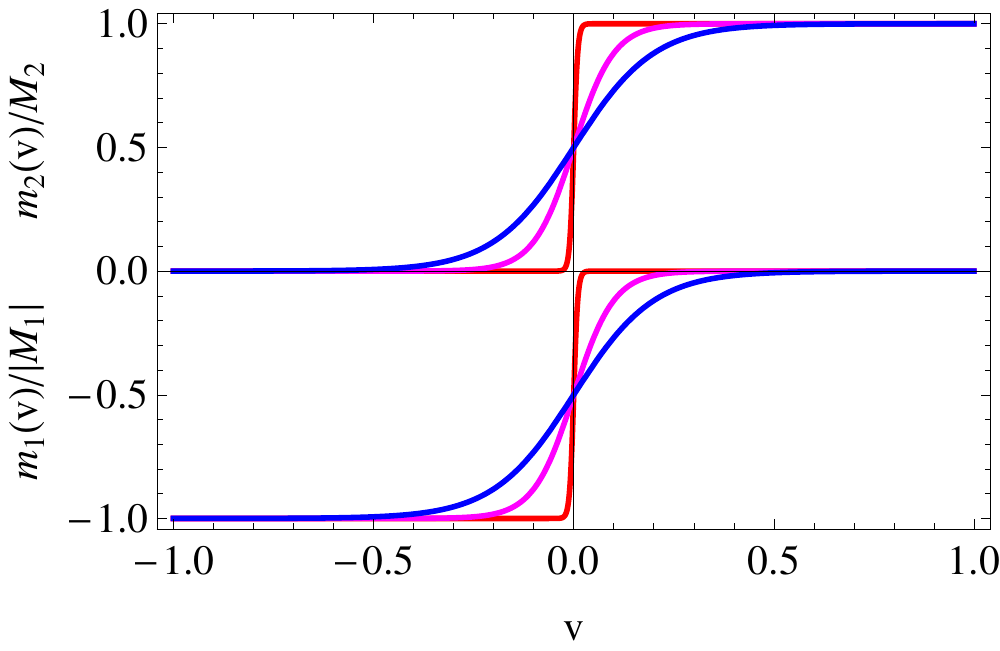}
\caption{\small Normalized mass functions $m_1(v)/|M_1|$ and $m_2(v)/M_2$ for various values of $v_0=0.01$ (red), $0.1$ (pink) and $0.2$ (blue).}
\label{m1m2}
\end{center}
\end{figure}

In Figure \ref{m1m2} we plot the behavior of the mass functions $m_1(v)/|M_1|$ and $m_2(v)/M_2$ with respect to $v$ for various values of $v_0$. It is clear that in the thin shell limit, \emph{i.e.} as $v_0 \to 0$, the variation of the mass functions is sharply localized around $v=0$ and approximates to a step function. Also, we have to bear in mind that the value of $M_1$ will depend on the dimension $d$ according to (\ref{M1M2val}).

In the remainder of this section, we will study the evolution of entanglement entropy in these time-dependent backgrounds in the thin-shell approximation.

\subsection{Holographic thermalization}
We are now ready to compute the entanglement entropy in our dynamical background. For the ease of the computation, we will take as our entangling surface a spherical region of radius $R$ inside the static patch, so that $RH\leq1$. We define a rescaled radius, $\tilde{r}=rH$, and parametrize the extremal surface by functions $z(\tilde{r})$ and $v(\tilde{r})$. The area functional in this case is given by
\begin{equation}
\A=\Omega_{d-2}\int_0^{\tilde{R}}\frac{d\tilde{r}}{z^{d-1}}\frac{\tilde{r}^{d-2}}{(1-\tilde{r}^2)^{(d-2)/2}}\sqrt{\frac{1}{(1-\tilde{r}^2)^2}-f(z,v) v'^2-2v'z'}\,,
\end{equation}
with $\tilde{R}=R H\leq1$. Here the mass function is time-dependent and is given by either (\ref{m1fun}) or (\ref{m2fun}). The equations of motion resulting from the above area functional are quite involved and therefore we do not present them explicitly. Also note that we do not have any conservation equation since the action depends explicitly on $r$.

To solve these equations we impose the following boundary conditions
\begin{eqnarray} \label{bc}
z(\epsilon) = z_* + {\rm corrections}\ , \quad z'(\epsilon) = 0 + {\rm corrections}\ ,\\
v(\epsilon) = v_* +{\rm corrections}\ , \quad v'(\epsilon) = 0 + {\rm corrections}  \ ,
\end{eqnarray}
where $z_*$ and $v_*$ are two free parameters, and $\epsilon$ is a small number.  The ``corrections'' in the above expressions are obtained in the following way: we first write $z(\tilde{r})$ and $v(\tilde{r})$ as power series in $\tilde{r}$ and then we expand the equations of motion around $\tilde{r}=0$. Finally, we solve the resulting equations order by order and then we evaluate the solutions at $\tilde{r}=\epsilon$.

\begin{figure}[!htm]
\begin{center}
\includegraphics[angle=0,
width=0.65\textwidth]{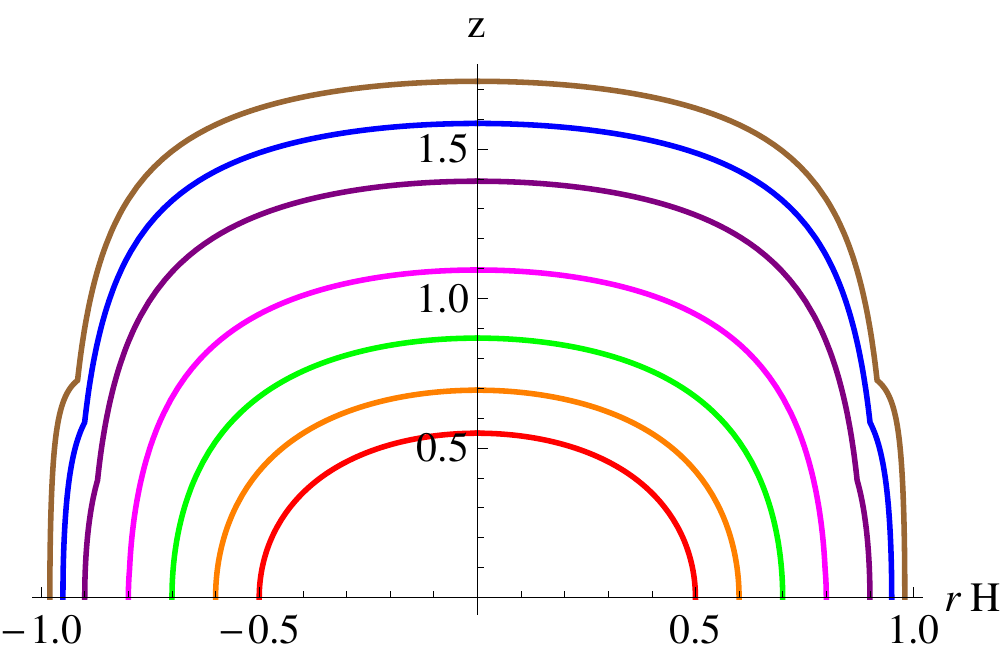}
\caption{\small Sample numerical solutions for $z(\tilde{r})$ in $d=2$. To obtain these plots we fixed $v_*=-1$ and chose various values of $z_*$. For this value of $v_*$, surfaces for which $RH\gtrsim0.9$ cross the shell and refract. For $d>2$ the behavior is qualitatively similar.}
\label{Geo}
\end{center}
\end{figure}

So far $z_*$ and $v_*$ are two free parameters that generate the numerical solutions for $z(\tilde{r})$ and $v(\tilde{r})$.
The boundary data can be obtained from
\begin{eqnarray}
z(\tilde{R}) = z_0 \ , \quad v (\tilde{R}) = \tilde{t} \ ,
\end{eqnarray}
where $z_0$ is an UV cut-off and $\tilde{t}\equiv tH$ is the boundary time. In Figure \ref{Geo} we plot sample numerical solutions for $z(\tilde{r})$ for fixed $v_*$ and various values of $z_*$. Some of them cross the shell located at $v=0$ and refract. For a fixed $\tilde{R}$ this always happens for arbitrary early times (arbitrary negative $v_*$). However, the converse is not always true. In flat space the extremal surfaces always cross the shell for arbitrary large volumes. However, in dS space we need $RH\leq1$ in order to have connected solutions. Hence, if we fix $v_*$, not always it is possible to find solutions that go deep enough into the bulk to cross the shell.

To generate the corresponding thermalization curves $S(t)$ for a fixed $\tilde{R}$, we do the following: for a fixed $v_*$ we vary $z_*$ until the read-off value of $\tilde{R}$ hits the desired value. This generates profiles for $z(\tilde{r})$ and $v(\tilde{x})$, which we can use to compute the area functional as well as the boundary time. Then, we vary $v_*$ and repeat the process again.

\begin{figure}[!htm]
$$
\begin{array}{cc}
  \includegraphics[angle=0,width=0.5\textwidth]{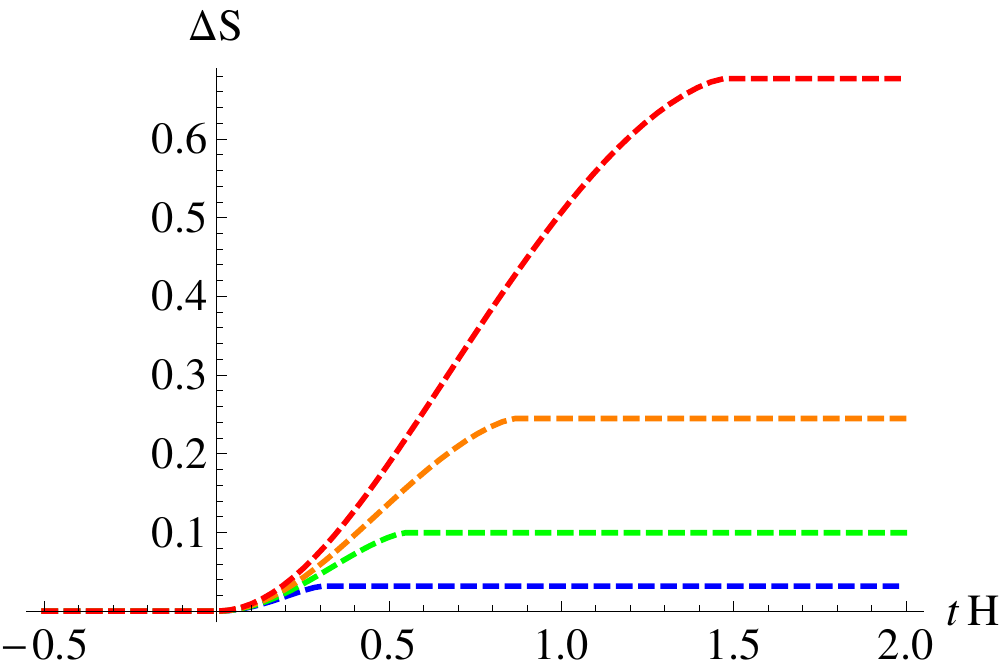} &\quad \includegraphics[angle=0,width=0.5\textwidth]{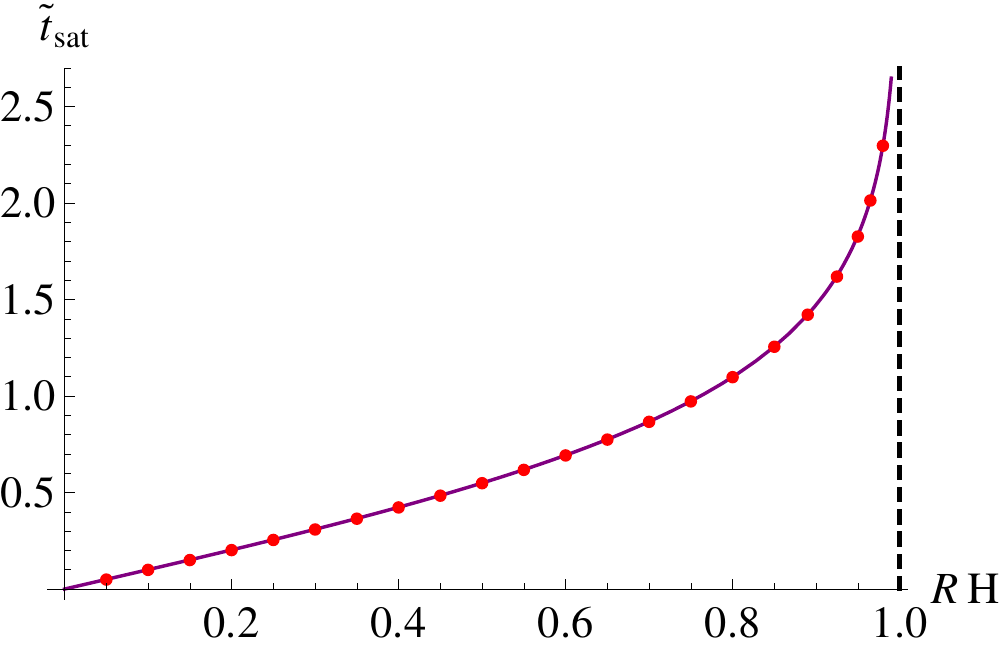}
\end{array}
$$
\caption{\small Left panel: evolution of $\Delta S$ in $d=2$ for different values of $RH=0.3$, 0.5, 0.7 and 0.9 from bottom to top. The situation is qualitatively similar for higher dimensions. Right panel: saturation time $\tilde{t}_{\mathrm{sat}}$ as a function of the sphere radius $RH$. Red dots correspond to numerical values for $\tilde{t}_{\mathrm{sat}}$ while the solid purple curve represent the fitting function $\tilde{t}_{\mathrm{sat}}=\tanh^{-1}(\tilde{R})$.}
\label{ther2m1}
\end{figure}

We are interested in the finite contribution to the entanglement entropy $S(t)$ for a fixed radius $\tilde{R}$. In order to study this quantity numerically, we fix the UV cutoff $\epsilon$ to be a small number, typically of the order of $10^{-3}$, and we define a rescaled entanglement entropy $\Delta S(t)$ by subtracting the entropy of the initial state $\Delta S(t)=S(t)-S(-\infty)$.\footnote{With this subtraction, the entanglement entropy $\Delta S(t)$ starts at zero in the infinite past.}\footnote{Another way to regularize would be by subtracting the entropy of a reference sphere of some fixed radius $\mathcal{R}$, $\Delta S_{\mathcal{R}}(t)=S_{\tilde{R}}(t)-S_{\mathcal{R}}(t)$. One can actually show that this quantity is finite, reflecting the fact that UV divergences do not depend on the value of $H$. However, time dependence varies with respect to the length-scale, so this quantity would not be useful in the present context.} Exploring the behavior of $\Delta S(t)$ as we change the radius $\tilde{R}$, we note some general properties. Qualitatively, our results agree with those of \cite{Liu:2013iza,Liu:2013qca} for Vaidya geometries with planar horizons. First, the evolution for very early times appears to be weakly dependent on the sphere size. The pre-local-equilibration regime is almost quadratic in time and it is followed by a post-local-equilibration linear growth phase at intermediate times. The entropy grows faster as $\tilde{R}$ is increased. Finally, there is a period of memory loss prior to equilibration and then the entropy abruptly flattens out at late times after it reaches saturation. In Figure \ref{ther2m1} (left panel) we plot this behavior for different values of the sphere radius. The saturation time $\tilde{t}_{\mathrm{sat}}$, on the other hand, shows a strong dependence on the sphere radius. It first increases linearly as $\tilde{R}$ is increased, with a slope of order unity, and then blows up logarithmically as $\tilde{R}$ approaches the horizon. This is surprising, it suggests the existence  of light-like degrees of freedom that do not random walk, in agreement with the results of \cite{Calabrese}. To see this, note that a radially outward light rays in de Sitter space obey the geodesic equation
\be
\frac{dr}{dt}=1-H^2r^2\,,
\ee
which can be solved and inverted to obtain $\tilde{t}=\tanh^{-1}(\tilde{r})$. For $\tilde{r}\ll1$ we get $\tilde{t}= \tilde{r}+\mathcal{O}(\tilde{r}^2)$, whereas for $\tilde{r}\sim1$ the leading contribution behaves like $\tilde{t}= \tfrac{1}{2} \left(\log(2)-\log(1-r)\right)+\mathcal{O}(1-\tilde{r})$. This result holds true for arbitrary dimension. We verified numerically this behavior for different values of $d=2,3,4$ and we found agreement to high accuracy. The results are plotted in Figure \ref{ther2m1} (right panel). It is important to mention that this behavior might be particular for spherical regions. If this holds true for more general situations, one might conjecture that the saturation time is given by the time it takes for a light-ray to reach the farthest point of a particular entangling surface with respect to the observer. However, to prove this conjecture one would need a microscopic description of the thermalization process, which is beyond the scope of this paper.

Finally, we also explored the possibility of varying the initial and final masses of the black hole, corresponding to quenches evolving from and to different thermal states. One of such examples was the mass function given by equation (\ref{m2fun}). While the equilibration value of $S_{\mathrm{sat}}$ showed a significant variation depending on the situation (and on the value of $d$), the behavior of $\tilde{t}_{\mathrm{sat}}$ was found to be a robust feature, and quite insensitive to the details of the quench.

\section{Conclusions and future directions\label{sec7}}

In this paper we have studied several properties of entanglement entropy in QFTs in de Sitter space, both in the static patch and the conformally flat patch. The theories in consideration have bulk duals coming from the standard Einstein-Hilbert action with negative cosmological constant and hence can be thought as models that belong to a universality class of strongly-coupled CFTs in the large-$N$ limit.

We started by choosing  the standard vacuum state  and obtained analytically extremal surfaces in the bulk with boundary condition taken as a spherical region of definite radius. According to the Ryu-Takayanagi prescription, the area of these solutions is interpreted as entanglement entropy for a spherical region in the boundary theory. Behaviors of extremal surfaces for $R<H^{-1}$ and $R>H^{-1}$ are qualitatively different. This implies   that the entanglement entropy and renormalized entanglement entropy undergo  {\it phase transitions} at $R=H^{-1}$. When $R<H^{-1}$, extremal surfaces are connected and U-shaped; whereas for  $R>H^{-1}$, extremal surfaces are disconnected. We believe this is a bulk refection of a drastic decrease in entanglement at distances $R>H^{-1}$ in the boundary theory.  For realistic cosmological scenarios where a period of accelerated expansion is followed by  decelerated expansion, regions outside the horizon will re-enter the horizon. It is reasonable to speculate that in that case, the extremal surfaces will  go through transitions from disconnected to connected shapes and consequently the entanglement entropy will also undergo a reverse phase transition. We will investigate this transition in more detail in the future.

We have also studied the rich phase structure of {\it entanglement/disentanglement phase-transition} of  mutual information in $(1+1)$ dimensions. We found that mutual information between any two intervals separated by a proper distance $x\ge 2/H$ is identically zero, implying that for any two intervals $A$ and $B$, separated by a distance larger than the horizon size, the density matrix $\rho_{A\cup B}=\rho_A \otimes\rho_B$  and hence they are completely disentangled. It is difficult to compute mutual information of spherical regions in higher dimensions, however, we expect qualitatively a similar behavior of mutual information in higher dimensions. A detailed calculation of mutual informations in higher dimensions would undoubtedly be useful to make this conclusion more robust.

We also considered out-of-equilibrium configurations in which the theory is undergoing a thermal quench. In order to do so, we constructed the Vaidya version of the well-known hyperbolic (or topological) black holes. In the static case, these solutions represent other states of the boundary theory (non-Bunch-Davies states) that are maintained externally in thermal equilibrium at a temperature $T\neq T_{\text{dS}}$. When time dependence is introduced, we claim that these bulk backgrounds describe the dynamics of dS QFTs in a thermalizing, out-of-equilibrium state. We gave particular attention to situations in which the initial state is taken to be at zero temperature and ends as a thermal bath at temperature $T=T_{\text{dS}}$, but we showed that our results hold for more general situations. For a fixed spherical region, the entanglement entropy follows a series of stages similar to what was found in \cite{Liu:2013iza,Liu:2013qca} for theories living in flat space. The saturation time $t_{\text{sat}}$, on the other hand, is found to depend on the sphere radius $R$: for $RH\ll 1$ it increases linearly but then it blows up as $RH\to1$. This result accounts for the fact that, for a static observer, any signal that is sent radially outwards takes an infinitely amount of time to reach the horizon due to an infinite blueshift. The behavior of $t_{\text{sat}}$ is independent of the number of dimensions. We argue that this behavior could be accounted for if we think of the process of thermalization in terms of a streaming of light-like degrees of freedom.

Last but not least, it is worth emphasizing that the theories we are considering in the present paper are conformal field theories. In addition, we took  spherical regions for the entangling surfaces which preserve the symmetries of the static patch. It would be interesting to investigate entanglement entropy for other shapes of the entangling surface, although it would be computationally more challenging. Another interesting possibility to address in the future would be to investigate the behavior of entanglement entropy in non-conformal theories on de Sitter space (see for instance \cite{Buchel:2002wf,Buchel:2002kj,Buchel:2006em,Buchel:2013dla}) and study the interplay between $H$ and the different phase transitions.

\section*{Acknowledgements}
We are grateful to D. Marolf, G. Pimentel  and M. Rangamani for useful correspondences.
This material is based upon work supported by the National Science Foundation under Grant Numbers
PHY-1316033 and PHY-0969020, and by Texas Cosmology Center, which is supported
by the College of Natural Sciences and the Department of Astronomy at the University
of Texas at Austin and the McDonald Observatory.

\end{document}